\documentclass[onecolumn,showpacs]{revtex4}

\usepackage{graphicx}
\usepackage{dcolumn}
\usepackage{bm}

\input epsf

\begin{document}
\title{\Large Higher Dimensional Cosmology with Some Dark Energy Models in Emergent,
Logamediate and Intermediate Scenarios of the Universe}

\author{\bf Chayan Ranjit$^1$\footnote{chayanranjit@gmail.com} Shuvendu Chakraborty$^1$\footnote{shuvendu.chakraborty@gmail.com} and Ujjal
Debnath$^2$\footnote{ujjaldebnath@yahoo.com,
ujjal@iucaa.ernet.in}}

\affiliation{$^1$Department of Mathematics, Seacom Engineering College, Howrah - 711 302, India.\\
$^2$Department of Mathematics, Bengal Engineering and Science
University, Shibpur, Howrah-711 103, India.}

\date{\today}

\begin{abstract}
We have considered $N$-dimensional Einstein field equations in
which four-dimensional space-time is described by a FRW metric and
that of extra dimensions by an Euclidean metric. We have chosen
the exponential forms of scale factors $a$ and $d$ numbers of $b$
in such a way that there is no singularity for evolution of the
higher dimensional Universe. We have supposed that the Universe is
filled with K-essence, Tachyonic, Normal Scalar Field and
DBI-essence. Here we have found the nature of potential of
different scalar field and graphically analyzed the potentials and
the fields for three scenario namely Emergent Scenario,
Logamediate Scenario and Intermediate Scenario. Also graphically
we have depicted the geometrical parameters named
\textit{statefinder parameters} and \textit{slow-roll parameters}
in the higher dimensional cosmology with the above mentioned
scenarios.
\end{abstract}


\maketitle

\section{\normalsize\bf{Introduction}}

From recent observations it is strongly believed that the most
interesting problems of particle physics cosmology are
  the origin due to accelerated expansion of the present Universe.
   The observation from type Ia supernovae [1,2] in associated with
   Large scale Structure [3] and Cosmic Microwave Background
    anisotropies(CMB) [4] have shown the evidences to support
    cosmic acceleration. The theory of Dark energy is the main
    responsible candidate for this scenario. From recent cosmological observations
    including supernova data [5] and measurements of cosmic microwave
    background radiation(CMBR) [4] it is evident that our present Universe
     is made up of about 4\% ordinary matter, about 74\% dark
      energy and about 22\% dark matter. Several interesting mechanisms
      have been suggested to explain this feature of this Universe,
       such as Loop Quantum Cosmology (LQC) [6], modified gravity [7],
        Higher dimensional phenomena [8], Brans-Dicke theory [9],  brane-world model [10] and many
        others.\\

Recently many cosmological models have been constructed by
introducing dark energies such as Phantom [11], Tachyon scalar
field [12], Hessence [13], Dilaton scalar field [14], K-essence
scalar field [15], DBI essence scalar field [16], and many others.
After realizing that many interesting of particle interactions
need more than four dimensions for their formulation, the study of
higher dimensional theory has been revived. The model of higher
dimensions was proposed by Kaluza and Klein [17,18] who tried to
introducing an extra dimension which is basically an extension of
Einstein general relativity in 5D. The activities of extra
dimensions also verified from the STM theory [19] proposed
recently by Wesson et al [20].  As our space-time is explicitly
four dimensional in nature so the `hidden' dimensions must be
related to the dark matter and dark energy which are also
`invisible' in nature.\\

Form the cosmological observation the present phase of
acceleration of the Universe is not clearly understood. Standard
Big Bang cosmology with perfect fluid assumption fails to
accommodate the observational fact. Recently, Ellis and Maartens
[21] have considered a cosmological model where inflationary
cosmologies exist in which the horizon problem is solved before
inflation begins, no big-bang singularity exist, no exotic physics
is involved and quantum gravity regime can even be avoided. An
emergent Universe model if developed in a consistent way is
capable of solving the conceptual problems of the big-bang model.
Actually the Universe starts out in the infinite past as an almost
static Universe and expands slowly, eventually evolving into a hot
big-bang era. An interesting example of this scenario is given by
Ellis, Murugan and Tsagas [22], for a closed Universe model with a
minimally coupled scalar field $\phi$, which has a special form of
interaction potential $V(\phi)$. There are several features for
the emergent Universe [21,23] viz. (i) the Universe is almost
static at the finite past, (ii) there is no time like singularity,
(iii) the Universe is always large enough so that the classical
description of space time is adequate, (iv) the Universe may
contains exotic matter so that the energy condition may be
violated, (v) the Universe is accelerating etc.\\

Here we also consider another two scenarios: (i) ``intermediate
scenario" and (ii) ``logamediate scenario" [24-27] to study of the
expanding anisotropic Universe in the presence of different scalar
fields. In the first case the scale factors evolves separately as
$a(t)=\exp(At^{f_{1}})$ and $b(t)=\exp(Bt^{f_{2}})$ where $A >0$,
$ B > 0$, $ 0 < f_{1} < 1$ and $ 0 < f_{2} < 1$. So the expansion
of the Universe is slower than standard de Sitter inflation
(arises when $f_{1} = f_{2} = 1$) but faster than power law
inflation with power greater than 1. The Harrison - Zeldovich
spectrum of fluctuation arises when $f_{1} = f_{2} = 1$ and $f_{1}
= f_{2} = 2/3$. In the second case we analyze the inflation with
scale factors separately of the form $a(t)=\exp(A(\ln
t)^{\lambda_{1}})$ and $b(t)=\exp(B(\ln t)^{\lambda _{2}})$ with
$A > 0$, $ B
> 0$, $\lambda _{1} > 1$ and $\lambda_{2} > 1$. When
$\lambda_{1}=\lambda _{2}=1$ this model reduces to power law
inflation. The logamediate inflationary form is motivated by
considering a class of possible cosmological solutions with
indefinite expansion which result from imposing
weak general conditions on the cosmological model.\\

In this work, we have considered N-dimensional Einstein field
equations in which 4-dimensional space-time is described by a FRW
metric and that of the extra $d$-dimensions by an Euclidean
metric. We also consider the Universe is filled with K-essence
scalar field, normal scalar field, tachyonic field and DBI essence
and investigate the natures of the dark energy candidates for
Emergent, Intermediate and Logamediate scenarios of the Universe.
Here in extra dimensional phenomenon we have shown the change of
the potential $V(\phi)$ corresponding to the field $\phi$ for the
dark energies mentioned above and also analyze  the anisotropic
Universe using the ``{\it slow roll}' parameters in
Hamilton-Jacobi formalism and in terms of above mentioned scalar
field $\phi$ and they are given by [24]

\begin{equation}
\epsilon = \frac{2\dot{H}^{2}}{H^{2}\dot{\phi}^{2}} \quad
\text{and}\quad
\eta=\frac{2}{H}\left[\frac{\dot{\phi}\ddot{H}-\dot{H}\ddot{\phi}}{\dot{\phi}^{3}}\right]
\end{equation}

Sahni et al [28] proposed the trajectories in the \{$r,s$\} plane
corresponding to different cosmological models to depict
qualitatively different behavior. The statefinder diagnostic along
with future SNAP observations may perhaps be used to discriminate
between different dark energy models. The above statefinder
diagnostic pair for higher dimensional anisotropic cosmology are
constructed from the scale factors $a(t)$ and $b(t)$ as follows:

\begin{equation}
r=1+3\frac{\dot{H}}{H^{2}}+\frac{\ddot{H}}{H^{3}} ~~\text{and} ~~
s=\frac{r-1}{3(q-\frac{1}{2})}
\end{equation}
where $q$ is the deceleration parameter defined by
$q=-1-\frac{\dot{H}}{H^{2}}$ and $H$ is the Hubble parameter.
Since this parameters are dimensionless so they allow us to
characterize the properties of dark energy in a model
independently. Finally we graphically analyzed geometrical
parameters ${r, s}$ in the higher dimensional anisotropic Universe
in emergent, logamediate and intermediate scenarios of the universe.\\

\section{\normalsize\bf{Basic Equations}}

We consider homogeneous and anisotropic $N$-dimensional space-time
model described by the line element [29,30]

\begin{equation}
ds^{2}=ds^{2}_{FRW}+\sum_{i=1}^{d}b^{2}(t)dx_{i}^{2}
\end{equation}

where $d$ is the number of extra dimensions $(d=N-4)$ and
$ds^{2}_{FRW}$ represents the line element of the FRW metric in
four dimensions is given by

\begin{equation}
ds^{2}_{F R
W}=-dt^{2}+a^{2}(t)\left[\frac{dr^{2}}{1-kr^{2}}+r^{2}(d\theta^{2}+\sin^{2}\theta
d\phi^{2})\right]
\end{equation}

where $a(t)$ and $b(t)$ are the functions of $t$ alone represent
the scale factors of 4-dimensional space time and extra
$d$-dimensions respectively. Here  $k ~(=0, ~\pm 1)$  is  the
curvature  index of the corresponding 3-space, so  that  the above
Universe is
described  as  flat, closed  and  open respectively.\\

The Einstein's field equations for the above non-vacuum higher
dimensional space-time symmetry are

\begin{equation}
3\left(\frac{\dot{a}^{2}+k}{a^{2}}\right)=\frac{\ddot{D}}{D}-\frac{d^{2}}{8}
\frac{\dot{b}^{2}}{b^{2}}+\frac{d}{8}
\frac{\dot{b}^{2}}{b^{2}}+\rho
\end{equation}

\begin{equation}
2\frac{\ddot{a}}{a}+\frac{\dot{a}^{2}+k}{a^{2}}=\frac{\dot{a}}{a}\frac{\dot{D}}{D}+\frac{d^{2}}{8}
\frac{\dot{b}^{2}}{b^{2}}-\frac{d}{8} \frac{\dot{b}^{2}}{b^{2}}-p
\end{equation}
and
\begin{equation}
\frac{\ddot{b}}{b}+3\frac{\dot{a}}{a}\frac{\dot{b}}{b}+\frac{\dot{D}}{D}\frac{\dot{b}}{b}-\frac{\dot{b}^{2}}{b^{2}}=-\frac{p}{2}
\end{equation}

where $\rho$ and $p$ are energy density and isotropic pressure
respectively. Here we choose here $8 \pi G=c=1$ and
$D^{2}=b^{d}(t)$, so we have
$\frac{\dot{D}}{D}=\frac{d}{2}\frac{\dot{b}}{b}$ and
$\frac{\ddot{D}}{D}=\frac{d}{2}\frac{\ddot{b}}{b}+\frac{d^{2}-2d}{4}\frac{\dot{b}^{2}}{b^{2}}$.
Also in this model we define the Hubble parameter as
$H=\frac{1}{d+3}(3\frac{\dot{a}}{a}+d\frac{\dot{b}}{b})$~.\\

\section{\normalsize\bf{Emergent Scenario}}

At first we consider Emergent scenario, where the scale factors
$a(t)$ and $b(t)$ are consider as the power of cosmic time $t$ are
given by [23, 31, 32]

\begin{equation}
a = a_{0}{(\beta + e^{\alpha t})}^{m} \quad and  \quad b =
b_{0}{(\mu + e^{\nu t})}^{n}
\end{equation}

 where $a_{0}$, $b_{0}$, $\alpha$, $\beta$, $\mu$, $\nu$, $m$ and $n$ are positive constants. So
the field equations (5), (6) and (7) become

\begin{equation}
\frac{3 m^{2} \alpha^{2} e^{2 \alpha t}}{(\beta + e^{\alpha
t})^{2}} + \frac{3 k}{a_{0}^{2}} (\beta + e^{\alpha t})^{-2
m}=\frac{d n \nu^{2} e^{2 \nu t}}{8 (\mu + e^{\nu t})^{2}}
(dn+n+4\mu e^{-\nu t})+\rho
\end{equation}

\begin{equation}
\frac{m \alpha^{2}e^{2 \alpha t}}{(\beta + e^{\alpha
t})^{2}}(3m+2\beta e^{-\alpha t})+\frac{k}{a_{0}^{2}} (\beta +
e^{\alpha t})^{-2 m}=\frac{d m n \alpha \nu e^{(\alpha + \nu)
t}}{2 (\beta + e^{\alpha t})(\mu + e^{\nu t})}+ \frac{d(d-1)n^{2}
\nu^{2} e^{2 \nu t}}{8(\mu + e^{2 \nu t})^{2}}-p
\end{equation}
and
\begin{equation}
\frac{n \mu \nu^{2} e^{\nu t}}{(\mu + e^{\nu t})^{2}}+\frac{3 m
n\alpha\nu e^{(\alpha+\nu)t}}{(\beta + e^{\alpha t})(\mu + e^{\nu
t})}+\frac{d n^{2}\nu^{2} e^{2\nu t}}{2(\mu + e^{\nu
t})^{2}}+\frac{p}{2}=0
\end{equation}

We now consider K-essence field, Tachyonic field,  normal scalar
field and DBI essence field. For these four cases we analyze the
behavior of the Emergent Universe in extra dimension and finally
we analyze the behavior of the statefinder parameters $r$ and $s$.\\

$\bullet$ \textbf{K-essence Field:}\\

The energy density and pressure due to K-essence field $\phi$ are
given by [15]

\begin{equation}
\rho= V(\phi)(-\chi+\chi^{2})
\end{equation}
and
\begin{equation}
p=V(\phi)(-\chi+3\chi^{2})
\end{equation}

where $\chi=\frac{\dot{\phi}^{2}}{2}$ and $V(\phi)$ is the relevant potential for K-essence Scalar field $\phi$.\\

Using equations (5)-(7), we can find the expressions for $V(\phi)$
and $\phi$ as

\begin{equation}
V(\phi) = \frac{\left(-\frac{48 m^{2}\alpha^{2}e^{2\alpha
t}}{(\beta + e^{\alpha t})^{2}}-\frac{24 m \alpha^{2}\beta
e^{\alpha t}}{(\beta + e^{\alpha t})^{2}}+\frac{6d m n \alpha \nu
e^{(\alpha+\nu)t}}{(\beta + e^{\alpha t})(\mu + e^{\nu
t})}+\frac{d(-1+2d)n^{2}\nu^{2}e^{2\nu t}}{(\mu + e^{\nu
t})^{2}}+\frac{2dn\mu \nu^{2}e^{\nu t}}{(\mu + e^{\nu
t})^{2}}-\frac{24k(\beta+e^{\alpha
t})^{-2m}}{a_{0}^{2}}\right)^{2}}{16\left(-\frac{12
m^{2}\alpha^{2}e^{2\alpha t}}{(\beta + e^{\alpha t})^{2}}-\frac{4
m \alpha^{2}\beta e^{\alpha t}}{(\beta + e^{\alpha t})^{2}}
+\frac{d m n \alpha \nu e^{(\alpha+\nu)t}}{(\beta + e^{\alpha
t})(\mu + e^{\nu t})}+\frac{d^{2} n^{2}\nu^{2}e^{2\nu t}}{2(\mu +
e^{\nu t})^{2}}+\frac{dn\mu \nu^{2}e^{\nu t}}{(\mu + e^{\nu
t})^{2}}-\frac{8k(\beta+e^{\alpha t})^{-2m}}{a_{0}^{2}}\right)}
\end{equation}
and
\begin{equation}
\phi=\int{2\left(\frac{-\frac{12 m^{2}\alpha^{2}e^{2\alpha
t}}{(\beta + e^{\alpha t})^{2}}-\frac{4 m \alpha^{2}\beta
e^{\alpha t}}{(\beta + e^{\alpha t})^{2}} +\frac{d m n \alpha \nu
e^{(\alpha+\nu)t}}{(\beta + e^{\alpha t})(\mu + e^{\nu
t})}+\frac{d^{2} n^{2}\nu^{2}e^{2\nu t}}{2(\mu + e^{\nu
t})^{2}}+\frac{dn\mu \nu^{2}e^{\nu t}}{(\mu + e^{\nu
t})^{2}}-\frac{8k(\beta+e^{\alpha t})^{-2m}}{a_{0}^{2}}}{-\frac{48
m^{2}\alpha^{2}e^{2\alpha t}}{(\beta + e^{\alpha t})^{2}}-\frac{24
m \alpha^{2}\beta e^{\alpha t}}{(\beta + e^{\alpha
t})^{2}}+\frac{6d m n \alpha \nu e^{(\alpha+\nu)t}}{(\beta +
e^{\alpha t})(\mu + e^{\nu t})}+\frac{d(-1+2d)n^{2}\nu^{2}e^{2\nu
t}}{(\mu + e^{\nu t})^{2}}+\frac{2dn\mu \nu^{2}e^{\nu t}}{(\mu +
e^{\nu t})^{2}}-\frac{24k(\beta+e^{\alpha
t})^{-2m}}{a_{0}^{2}}}\right)^{\frac{1}{2}}}~dt
\end{equation}

From above forms of $V$ and $\phi$, we see that $V$ can not be
expressed explicitly in terms of $\phi$.\\

$\bullet$ \textbf{Tachyonic  field:}\\

The energy density $\rho$ and pressure $p$ due to the Tachyonic
field $\phi$ are given by [12]

\begin{equation}
\rho=\frac{V(\phi)}{\sqrt{1- \dot{\phi}^{2}}}
\end{equation}
and
\begin{equation}
p=-V(\phi)\sqrt{1- \dot{\phi}^{2}}
\end{equation}
\begin{figure}

\includegraphics[scale=0.5]{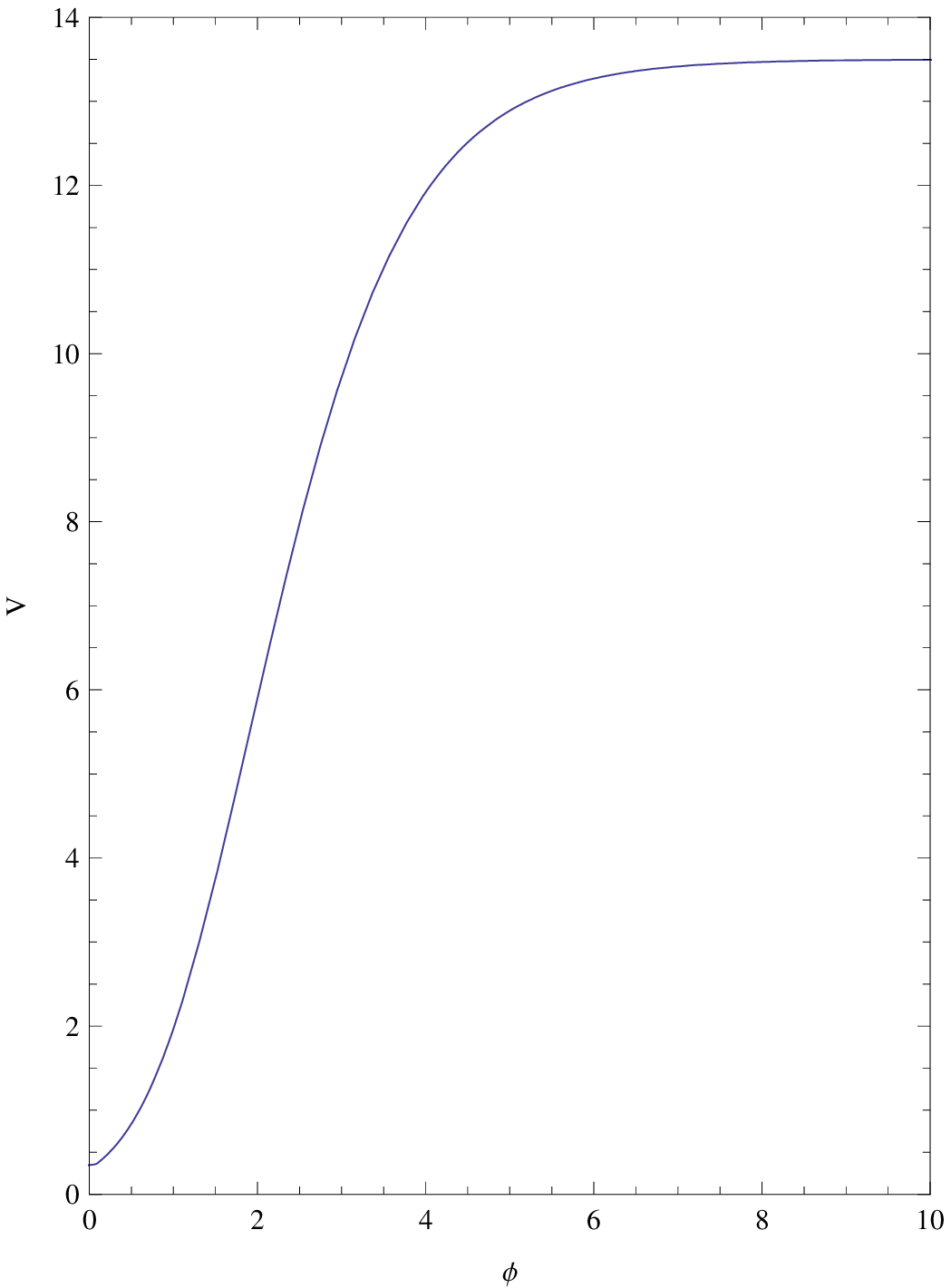}~~~~~~~~~
\includegraphics[scale=0.8]{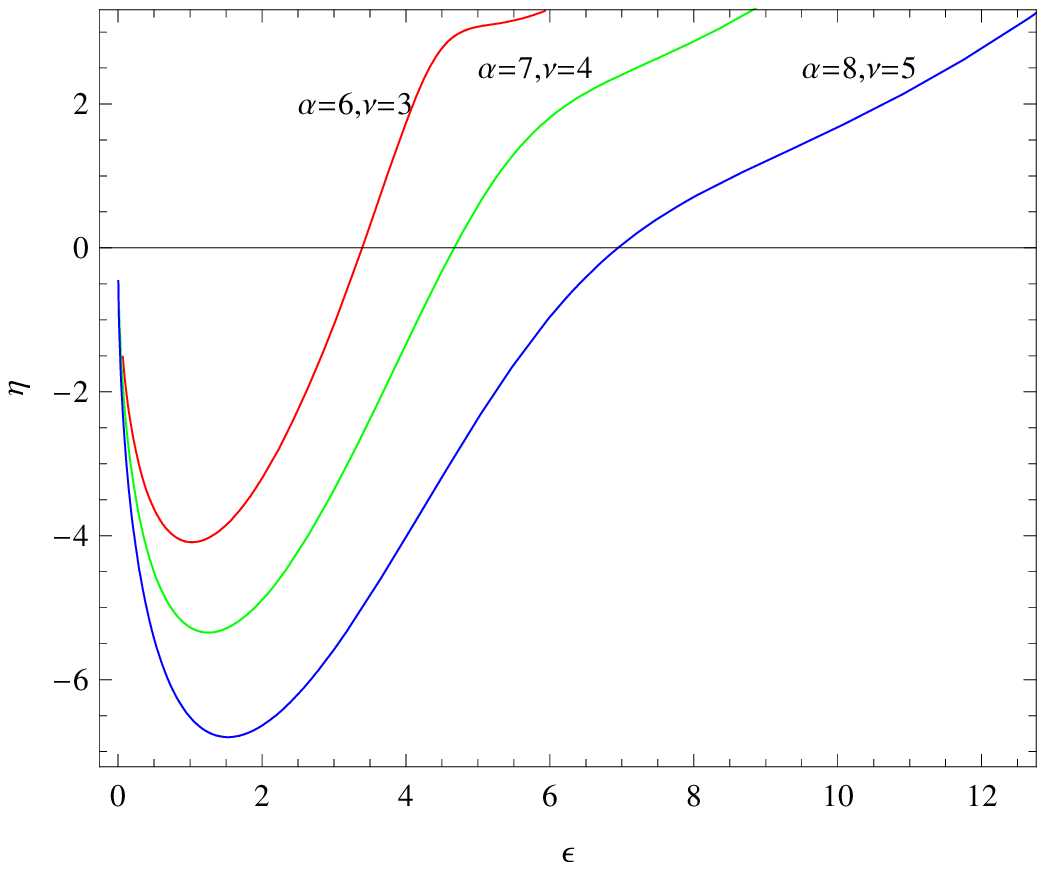}
\vspace{1mm}\\
~Fig.1~~~~~~~~~~~~~~~~~~~~~~~~~~~~~~~~~~~~~~~~~~~~~~~~~~~~~~~~~~~~~~~~~~~~~Fig.2~~~~~~\\

\vspace{6mm}  Fig. 1 shows the variations of $V$ against $\phi$,
for $a_{0} = 2,\alpha = 1.5,\beta = 1.6,\mu = 3,\nu = 1, m = 1, n
= 3, k = 1, d = 5$ and Fig. 2 shows the variation of the slow roll
parameters $\epsilon$ against $\eta$ for $m = 30, n = 10, \beta =
15, a_{0} = 1, \mu = 1, d = 40,$ $k= -1, 1, 0$ respectively in the
case of K-essence Scalar field Emergent Scenario.

 \vspace{6mm}

\end{figure}
where $V(\phi)$ is the relevant potential for the Tachyonic field
$\phi$. Using equations (5)-(7), we can find the expressions for
$V(\phi)$ and $\phi$ as

\begin{eqnarray*}
V(\phi) =\sqrt{\left(\frac{m \alpha^{2} e^{\alpha t}(2 \beta + 3m
e^{\alpha t})}{(\beta + e^{\alpha t})^{2}}- \frac{d m n \alpha \nu
e^{(\alpha+\nu)t}}{2(\beta + e^{\alpha t})(\mu + e^{\nu
t})}-\frac{d(d-1)n^{2}\nu^{2} e^{2\nu t}}{8(\mu + e^{\nu
t})^{2}}+\frac{k(\beta + e^{\alpha t})^{-2m}}{a_{0}^{2}}\right)}
\end{eqnarray*}

\begin{equation}
\times \sqrt{\left(\frac{3m^{2}\alpha^{2} e^{2\alpha t}}{(\beta +
e^{\alpha t})^{2}}-\frac{dn\nu^{2} e^{\nu t}(4\mu+(1+d)n e^{\nu
t})}{8(\mu + e^{\nu t})^{2}}+\frac{3k(\beta + e^{\alpha
t})^{-2m}}{a_{0}^{2}}\right)}
\end{equation}

and

\begin{equation}
\phi=\int{\sqrt{1+\frac{-\frac{8m \alpha^{2} e^{\alpha t}(2 \beta
+ 3m e^{\alpha t})}{(\beta + e^{\alpha t})^{2}}+ \frac{4d m n
\alpha \nu e^{(\alpha+\nu)t}}{(\beta + e^{\alpha t})(\mu + e^{\nu
t})}+\frac{d(d-1)n^{2}\nu^{2} e^{2\nu t}}{(\mu + e^{\nu
t})^{2}}-\frac{8k(\beta + e^{\alpha
t})^{-2m}}{a_{0}^{2}}}{8\left(\frac{3m^{2}\alpha^{2} e^{2\alpha
t}}{(\beta + e^{\alpha t})^{2}}-\frac{dn\nu^{2} e^{\nu
t}(4\mu+(1+d)n e^{\nu t})}{8(\mu + e^{\nu t})^{2}}+\frac{3k(\beta
+ e^{\alpha t})^{-2m}}{a_{0}^{2}}\right)}}}~dt
\end{equation}

From above forms of $V$ and $\phi$, we see that $V$ can not be
expressed explicitly in terms of $\phi$.\\\\

$\bullet$ \textbf{Normal Scalar field:}\\

The energy density $\rho$ and pressure $p$ due to the Normal
Scalar field $\phi$ are given by [37]

\begin{equation}
\rho=\frac{1}{2}\dot{\phi}^{2}+V(\phi)
\end{equation}
and
\begin{equation}
p=\frac{1}{2}\dot{\phi}^{2}-V(\phi)
\end{equation}

where $V(\phi)$ is the relevant potential for the Normal Scalar
field $\phi$. Using equations (5)-(7), we can find the expressions
for $V(\phi)$ and $\phi$ as

\begin{figure}
\includegraphics[scale=0.7]{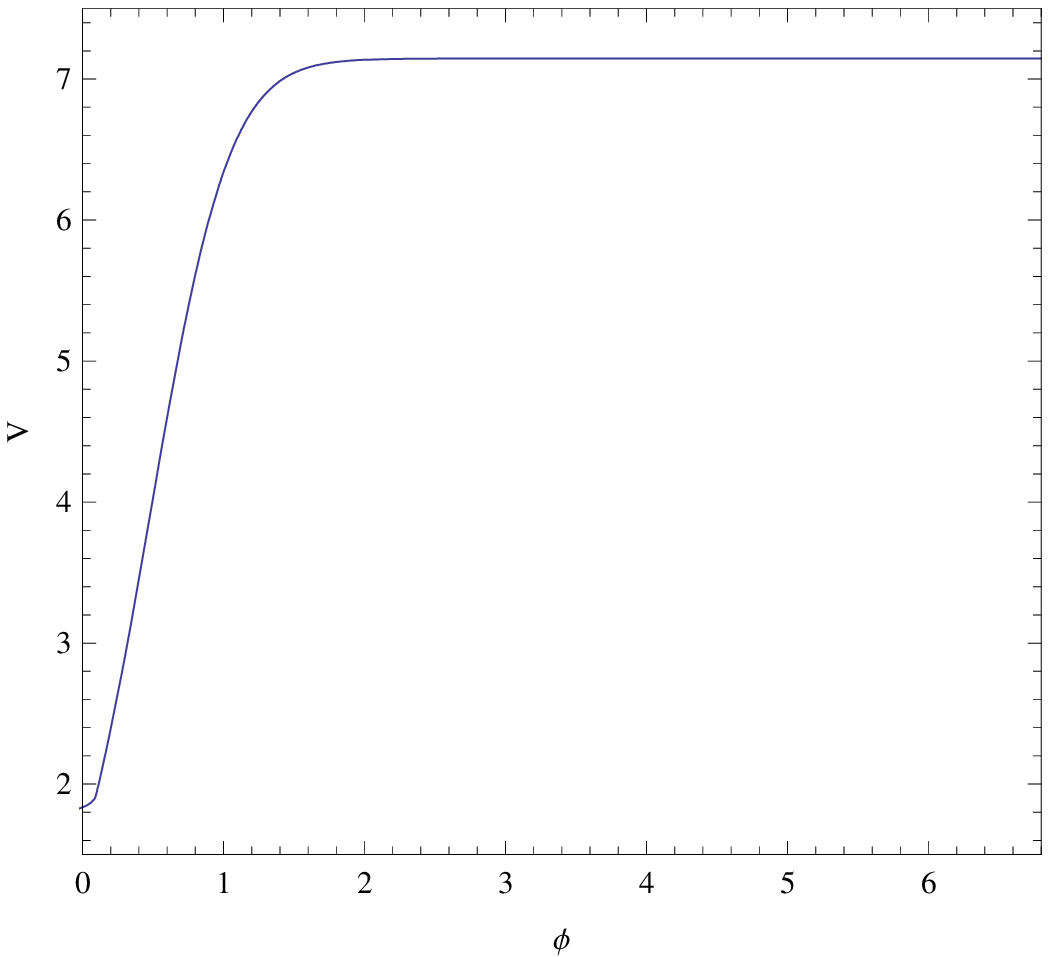}~~~~~~~~~~
\includegraphics[scale=0.77]{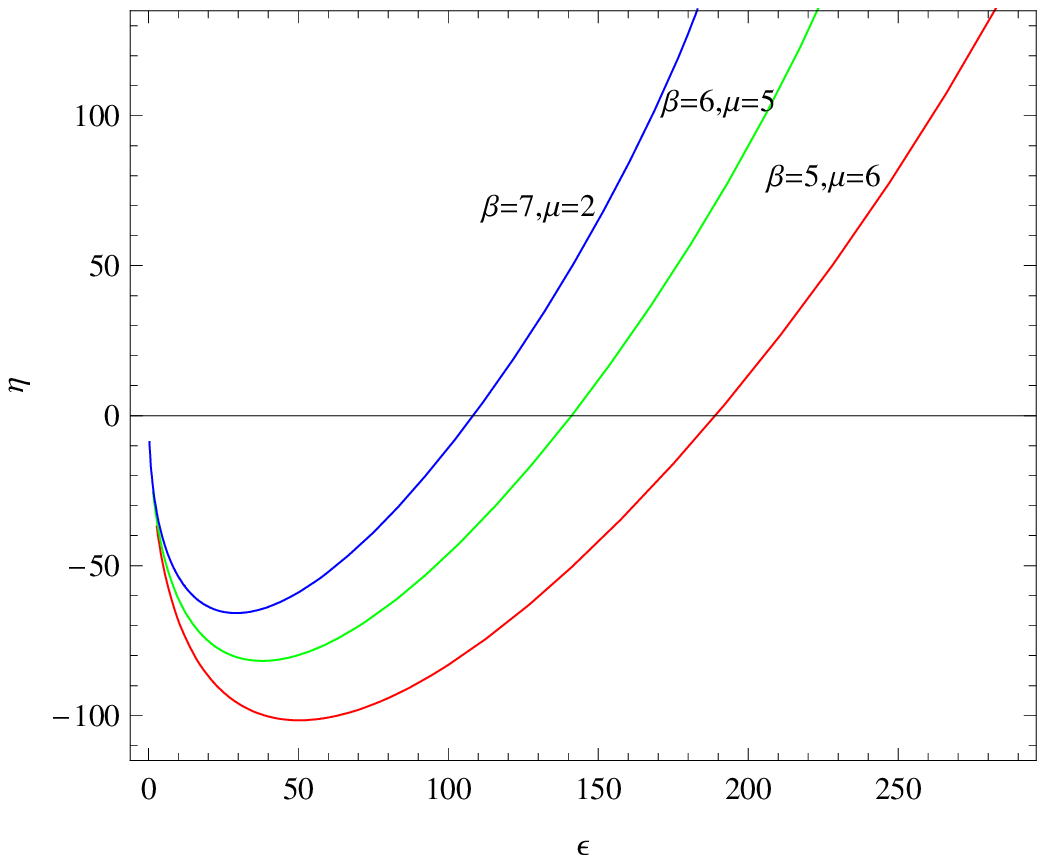}
\vspace{1mm}\\
~~~~~~~~~~~Fig.3~~~~~~~~~~~~~~~~~~~~~~~~~~~~~~~~~~~~~~~~~~~~~~~~~~~~~~~~~~~~~~~~~~~~~~~~~~~~~~~~Fig.4~~~~~~~~\\

\vspace{6mm} Fig. 3 shows the variations of $V$ against $\phi$,
for $a_{0} = 2,\alpha = 1.1,\beta = 1.5,\mu = 1.2,\nu = 1.5, m =
1, n = 2, k = 1, d = 5$ and Fig. 4 shows the variation of the slow
roll parameters $\epsilon$ against $\eta$ for $m = 30, n = 6,
\alpha=1 , a_{0} = 1, ,\nu = 5, d = 15,$ $k= -1, 1, 0$
respectively in the case of Tachyonic Scalar field for Emergent
Scenario.
 \vspace{6mm}

\end{figure}

\begin{equation}
V(\phi)=\frac{3m^{2}\alpha^{2}e^{2\alpha t}}{(\beta + e^{\alpha
t})^{2}}+\frac{m \alpha^{2}\beta e^{\alpha t}}{(\beta + e^{\alpha
t})^{2}}-\frac{d m n\alpha \nu e^{(\alpha+\nu)t}}{4(\beta +
e^{\alpha t})(\mu + e^{\nu t})}-\frac{d^{2}n^{2}\nu^{2}e^{2\nu
t}}{8(\mu + e^{\nu t})^{2}}-\frac{dn\mu \nu^{2}e^{\nu t}}{4(\mu +
e^{\nu t})^{2}}+\frac{2k(\beta+e^{\alpha t})^{-2m}}{a_{0}^{2}}
\end{equation}
and

\begin{equation}
\phi=\frac{1}{2}\int{\left(-\frac{8m \alpha^{2}\beta e^{\alpha
t}}{(\beta + e^{\alpha t})^{2}}+\frac{2d m n\alpha \nu
e^{(\alpha+\nu)t}}{(\beta + e^{\alpha t})(\mu + e^{\nu
t})}-\frac{d n^{2}\nu^{2}e^{2\nu t}}{(\mu + e^{\nu
t})^{2}}-\frac{2dn\mu \nu^{2}e^{\nu t}}{(\mu + e^{\nu
t})^{2}}+\frac{8k(\beta+e^{\alpha t})^{-2m}}{a_{0}^{2}}\right)}~dt
\end{equation}\\\\

$\bullet$ \textbf{DBI-essence:}\\

The energy density $\rho$ and pressure $p$ due to the DBI-essence
field $\phi$ are given by [35,36]

\begin{equation}
\rho=(\gamma-1)T(\phi)+V(\phi)
\end{equation}
and
\begin{equation}
p=\frac{\gamma-1}{\gamma}T(\phi)-V(\phi)
\end{equation}
where $\gamma$ is given by
\begin{equation}
\gamma=\frac{1}{\sqrt{1-\frac{\dot{\phi}^{2}}{T(\phi)}}}
\end{equation}
 and
$V(\phi)$ is the relevant potential for the DBI-essence field
$\phi$.

The energy conservation equation is given by
\begin{equation}
\dot{\rho}_{\phi}+3H(\rho_{\phi}+p_{\phi})=0
\end{equation}
where $H$ is the Hubble parameter in terms of scale factor as
\begin{equation}
H=\frac{1}{3}\left(3\frac{\dot{a}}{a}+d\frac{\dot{b}}{b}\right)
\end{equation}
From energy conservation equation we have the wave equation for
$\phi$ as

\begin{equation}
\ddot{\phi}-\frac{3T'(\phi)}{2T(\phi)}\dot{\phi}^{2}+T'(\phi)+\left(3\frac{\dot{a}}{a}+
d\frac{\dot{b}}{b}\right)\frac{\dot{\phi}}{\gamma^{2}}+
\frac{1}{\gamma^{3}}[V'(\phi)-T'(\phi)]=0
\end{equation}
where $'$ is the derivative with respect to $\phi$. Now for
simplicity of calculations, we consider two particular cases:
$\gamma$ = constant and $\gamma \neq$
constant.\\

{\bf Case I:}  $\gamma$ = constant.\\

In this case, for simplicity, we assume
$T(\phi)=\sigma\dot{\phi}^{2}$ $(\sigma> 1)$ and $V(\phi)=
\delta\dot{\phi}^{2}$ ($\delta>0$). So we have $\gamma =
\sqrt{\frac{\sigma}{\sigma-1}}$.\\
In these choices we have the following solutions for $V(\phi)$,
$T(\phi)$ and $\phi$ from equation (29)as

\begin{equation}
\phi=\int{\phi_{0}a_{0}^{F}b_{0}^{G}(\beta + e^{\alpha t})^{m F+n
G}}~dt
\end{equation}
 from
\begin{equation}
V(\phi)=\delta \phi_{0}^{2}a_{0}^{2F}b_{0}^{2G}(\beta + e^{\alpha
t})^{2m F+2n G}
\end{equation}

and

\begin{equation}
T(\phi)=\sigma\phi_{0}^{2}a_{0}^{2F}b_{0}^{2G}(\beta + e^{\alpha
t})^{2m F+2n G}
\end{equation}
where$E=\frac{1}{\gamma}[2(\delta-\sigma)+2(\sigma-1)\gamma^{3}]$,
$\gamma=\sqrt{\frac{\sigma}{\sigma-1}}$,$F=-\frac{3}{E}$,
$G=-\frac{d}{E}$ and $\phi_{0}$ is an integrating constants. From
Fig. 7 and Fig. 8, we see that $V(\phi)$ and $T(\phi)$ are both
exponentially decreasing with DBI scalar field
$\phi$.\\

\begin{figure}
\includegraphics[scale=0.7]{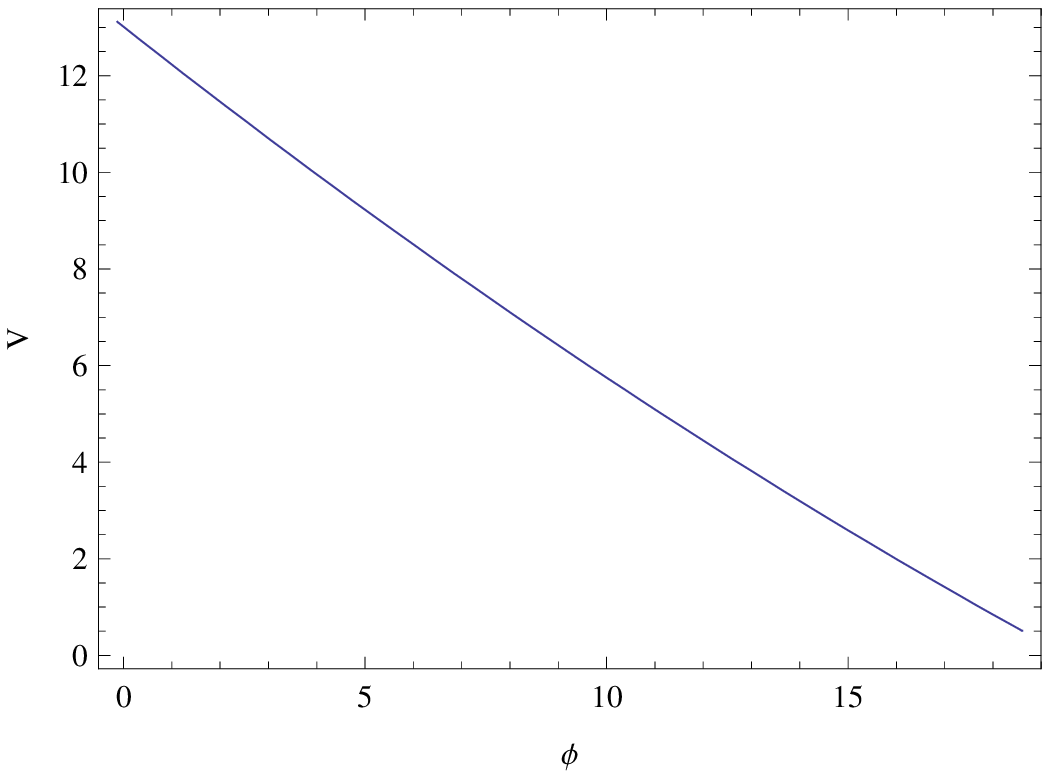}~~~~
\includegraphics[scale=.9]{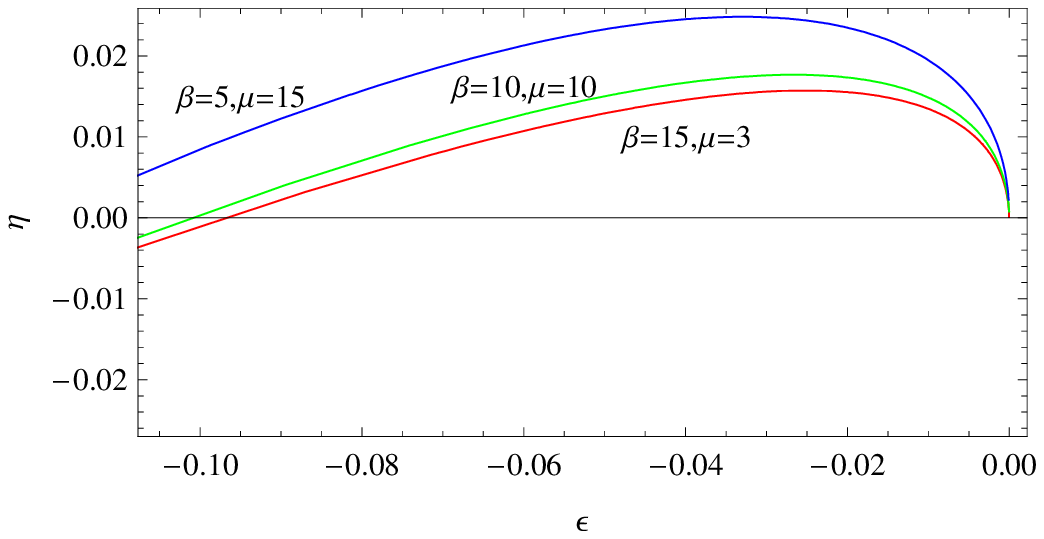}\\
\vspace{2mm}
~~~~Fig.5~~~~~~~~~~~~~~~~~~~~~~~~~~~~~~~~~~~~~~~~~~~~~~~~~~~~~~~~~~~~~~~~~~~~~~~~~~~~~~~~Fig.6\\
\vspace{6mm} Fig. 5 shows the variations of $V$ against $\phi$,
for $a_{0} =0.5 ,\alpha =0.5 ,\beta =0.5 ,\mu =0.1 ,\nu =0.1 , m
=1 , n =8 , k =1 , d =15 $ and Fig. 6 shows the variation of the
slow roll parameters $\epsilon$ against $\eta$ for $m =5 , n =4 ,
\alpha=3 , a_{0} =1 , ,\nu =7 , d =15 ,$ $k= -1, 1, 0$
respectively in the case of Normal Scalar field for Emergent
Scenario.
 \vspace{6mm}

\end{figure}
\begin{figure}
\includegraphics[scale=0.5]{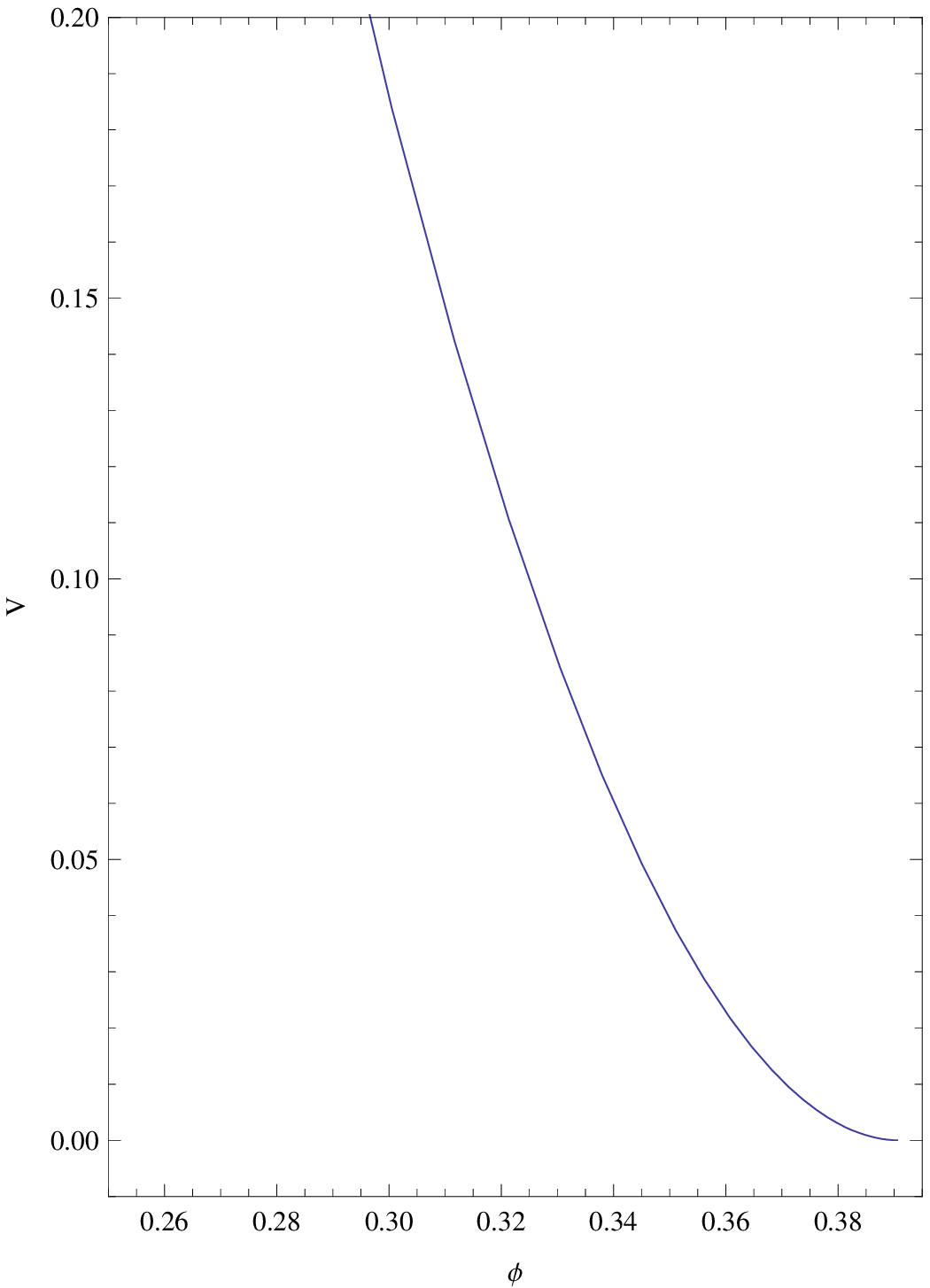}~~~~~~~~~~~~~~~
\includegraphics[scale=0.5]{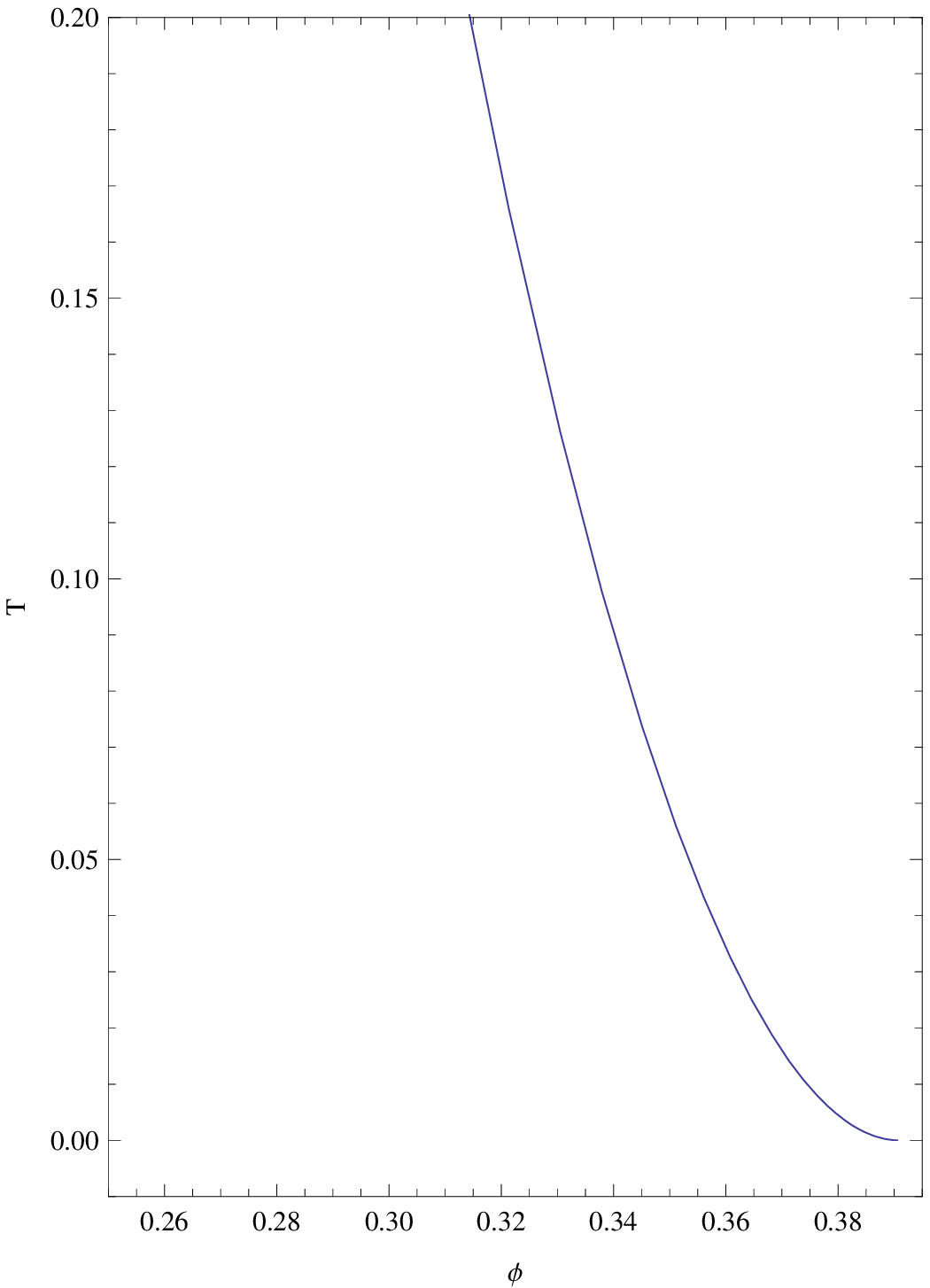}\\
\vspace{2mm}
~~~~~~~~Fig.7~~~~~~~~~~~~~~~~~~~~~~~~~~~~~~~~~~~~~~~~~~~~~~~~~~~~~~~~~~~~~~~Fig.8\\
\vspace{4mm}
\includegraphics[scale=.45]{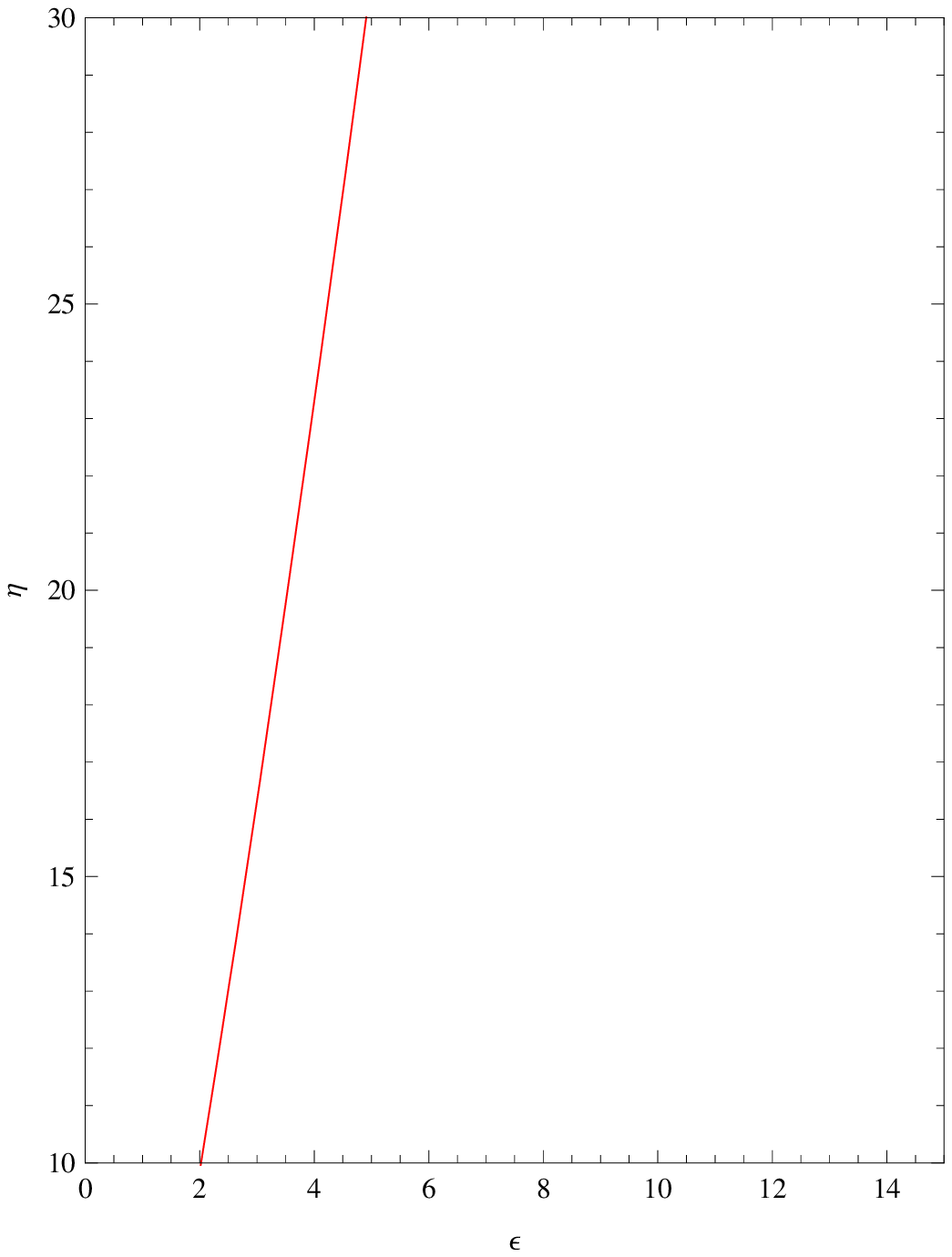}\\
\vspace{2mm}
~~~~~~~~~~~~~~~~~~~~~~~~~~~~~~~~~~~~Fig.9~~~~~~~~~~~~~~~~~~~~~~~~~~~~~~~~\\
\vspace{6mm} Fig. 7 shows the variations of $V$ against $\phi$,
and Fig. 8 shows the variations of $T$ against $\phi$ for $a_{0} =
0.2,b_{0}=.2,\alpha =0.5,\beta =0.6,m =15,n =3,k=1,d=15,\sigma=3,
\delta=2,\phi_{0}=2$ Fig. 9 shows the variation of the slow roll
parameters $\epsilon$ against $\eta$ for $a_{0}
=0.02,b_{0}=0.03,\alpha=0.7,\beta =0.8,m =2,n=5,d=5,
\sigma=2,\delta=5$ respectively in the $1st$ case of DBI-essence
Scalar field scenario.
 \vspace{6mm}
\end{figure}
\begin{figure}
\includegraphics[scale=0.45]{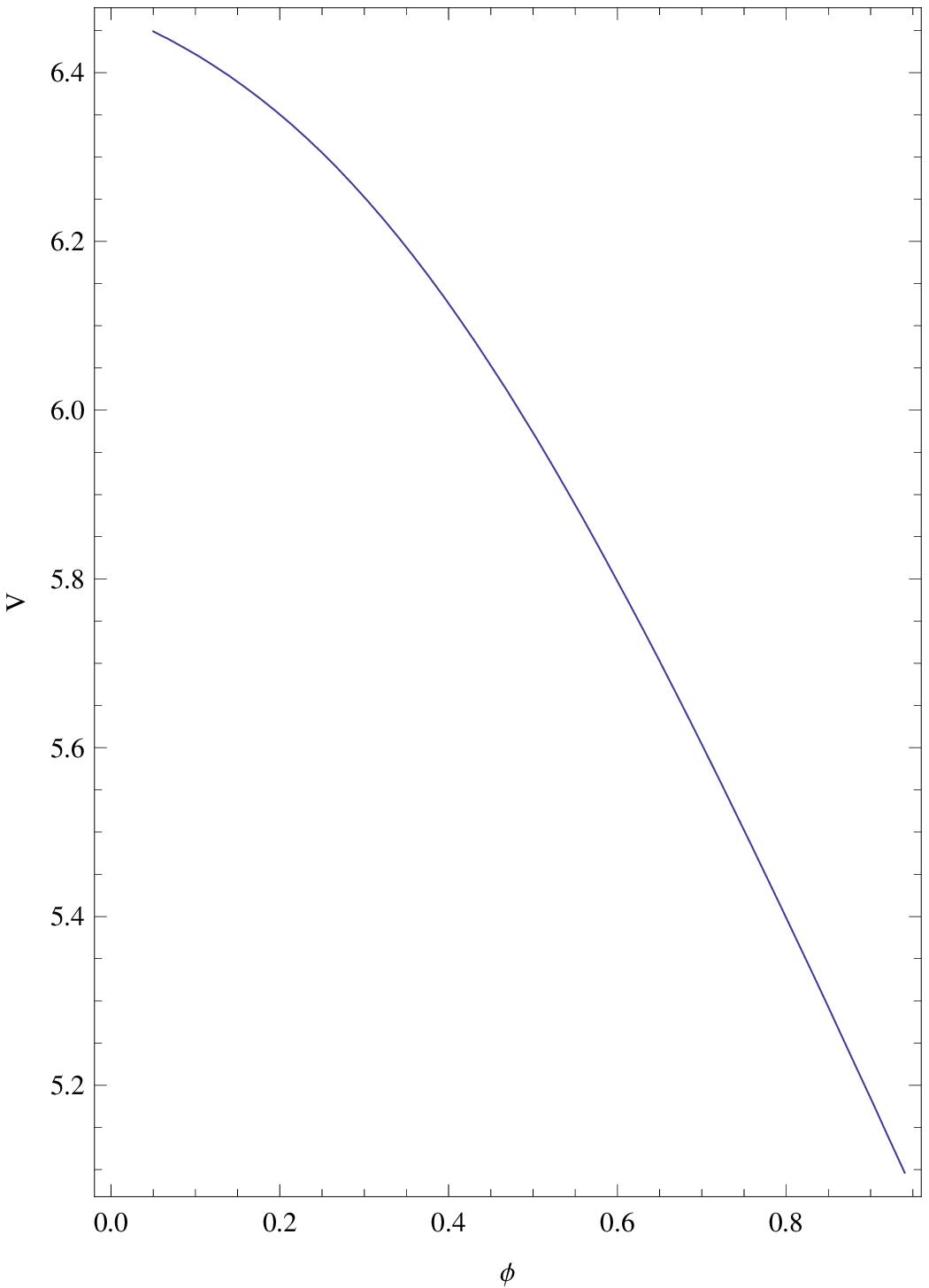}~~~~~~~~~~~~~~~~~~
\includegraphics[scale=.55]{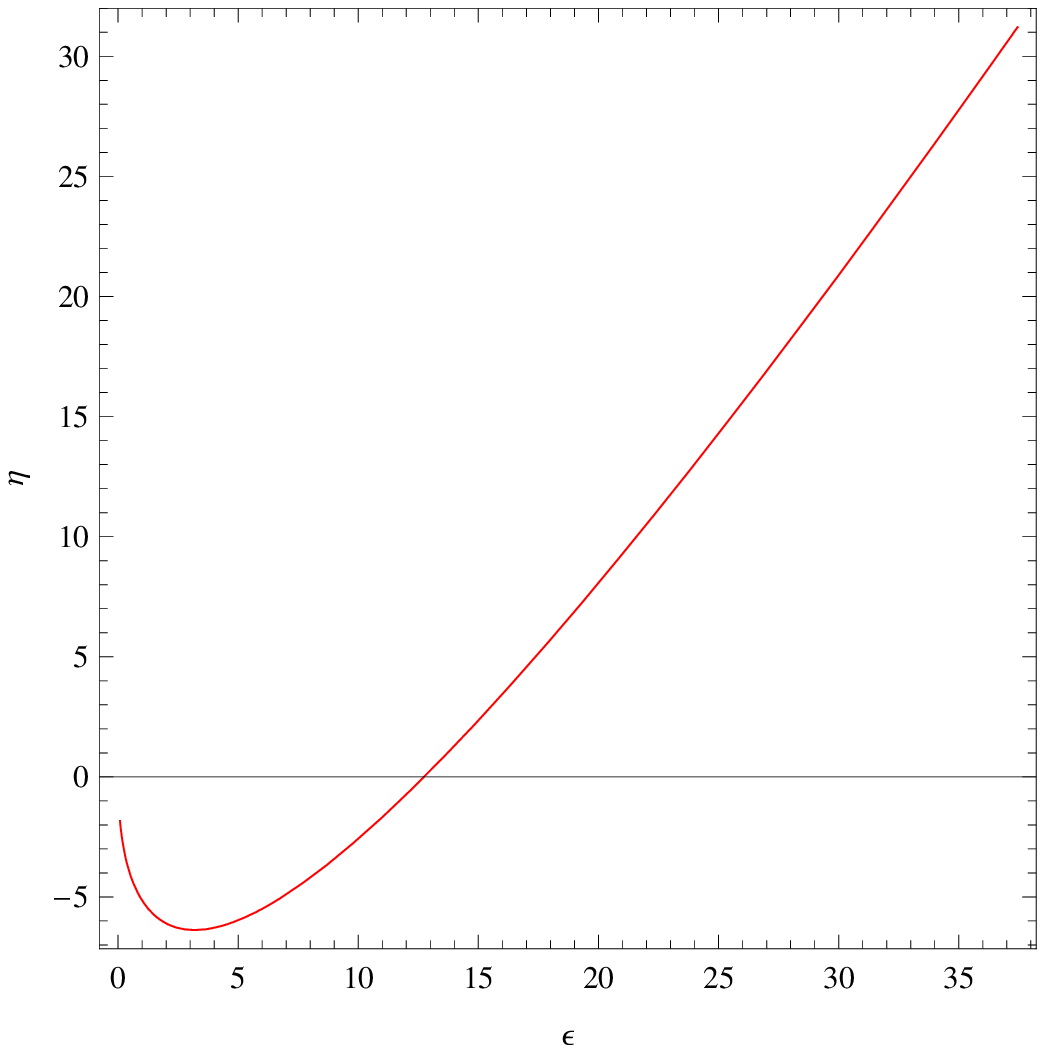}\\
\vspace{2mm}
~~~~Fig.10~~~~~~~~~~~~~~~~~~~~~~~~~~~~~~~~~~~~~~~~~~~~~~~~~~~~~~~~~~~~Fig.11~~~~\\
\vspace{6mm} Fig. 10 shows the variations of $V$ against $\phi$,
for$a_{0} =0.02,b_{0}=0.03,C_{0}=2,\alpha =5, \beta =6, m =0.1, n
=0.3,d =5 $ and Fig. 11 shows the variation of the slow roll
parameters $\epsilon$ against $\eta$ for $a_{0}=0.02, b_{0}=0.03,
C_{0}=2, \alpha=5, \beta=6,m=0.1,n=0.3,d =5$ respectively in the
$2nd$ case of DBI-essence Scalar field for Emergent Scenario.
 \vspace{6mm}
\end{figure}

{\bf Case II:}  $\gamma \neq$  constant.\\

In this case, we consider $\gamma=\dot{\phi}^{-2}$ and
$V(\phi)=T(\phi)$. Using equations (5)-(7) and (24)-(26), we can
find the expressions for $V(\phi)$, $T(\phi)$ and $\phi$ as [34]
\begin{equation}
V(\phi)=T(\phi)=\left(\ln{[\frac{C_{0}}{a_{0}^{3}b_{0}^{d}(\beta+e^{\alpha
t})^{(3m+dn)}}]}\right)\sqrt{\left(1-\frac{1}{\ln{[\frac{C_{0}}{a_{0}^{3}b_{0}^{d}(\beta+e^{\alpha
t})^{(3m+dn)}}]}}\right)}
\end{equation}

and
\begin{equation}
\phi=\int{\left(1-\frac{1}{\ln{[\frac{C_{0}}{a_{0}^{3}b_{0}^{d}(\beta+e^{\alpha
t})^{(3m+dn)}}]}}\right)^{\frac{1}{4}}}~dt
\end{equation}
where $C_{0}$ is an integrating constant.\\

$\bullet$ \textbf{Statefinder parameters:}\\

The geometrical parameters\{$r,s$\} for higher dimensional
anisotropic cosmology in emergent scenario can be constructed from
the scale factors $a(t)$ and $b(t)$ as

\begin{equation}
r=1+ \frac{(3+d)\left(3\left(\frac{3m\alpha e^{\alpha
t}}{\beta+e^{\alpha t}}+\frac{dn\nu e^{\nu t}}{\mu + e^{\nu
t}}\right)\left(\frac{3m\alpha^{2}\beta e^{\alpha
t}}{(\beta+e^{\alpha t})^{2}}+\frac{dn\mu \nu^{2}e^{\nu t}}{(\mu +
e^{\nu t})^{2}}\right) +(3+d)\left(\frac{3m\alpha^{3}\beta
e^{\alpha t}(\beta -e^{\alpha t})}{(\beta+e^{\alpha
t})^{3}}+\frac{dn\mu \nu^{3}e^{\nu t}(\mu -e^{\nu t})^{3}}{(\mu
+e^{\nu t})^{3}}\right)\right)}{(\frac{3m\alpha e^{\alpha
t}}{\beta+e^{\alpha t}}+\frac{dn\nu e^{\nu t}}{\mu + e^{\nu
t}})^{3}}
\end{equation}
\begin{equation}
s=\frac{\left((3+d)\left(3\left(\frac{3m\alpha e^{\alpha t}}{\beta
+e^{\alpha t}}+\frac{dn\nu e^{\nu t}}{\mu + e^{\nu
t}}\right)\left(\frac{3m\alpha^{2}\beta e^{\alpha t}}{(\beta+
e^{\alpha t})^{2}}+\frac{dn\mu \nu^{2}e^{\nu t}}{(\mu+ e^{\nu
t})^{2}}\right)+(3+d)\left(\frac{3m\alpha^{3}\beta e^{\alpha
t}(\beta -e^{\alpha t})}{(\beta+e^{\alpha t})^{3}}+\frac{dn\mu
\nu^{3}e^{\nu t}(\mu -e^{\nu t})}{(\mu +e^{\nu
t})^{3}}\right)\right)\right)}{\left(3\left(\frac{3m\alpha
e^{\alpha t}}{\beta+e^{\alpha t}}+\frac{dn\nu e^{\nu t}}{\mu +
e^{\nu t}}\right)^{3}\left(-\frac{3}{2}
-\frac{3(3+d)\left(\frac{3m\alpha^{2}\beta e^{\alpha t}}{(\beta+
e^{\alpha t})^{2}}+\frac{dn\mu \nu^{2}e^{\nu t}}{(\mu+ e^{\nu
t})^{2}}\right)}{\left(\frac{3m\alpha e^{\alpha
t}}{\beta+e^{\alpha t}}+\frac{dn\nu e^{\nu t}}{\mu + e^{\nu
t}}\right)^{2}}\right)\right)}
\end{equation}\\\\

\begin{figure}
\includegraphics[scale=0.8]{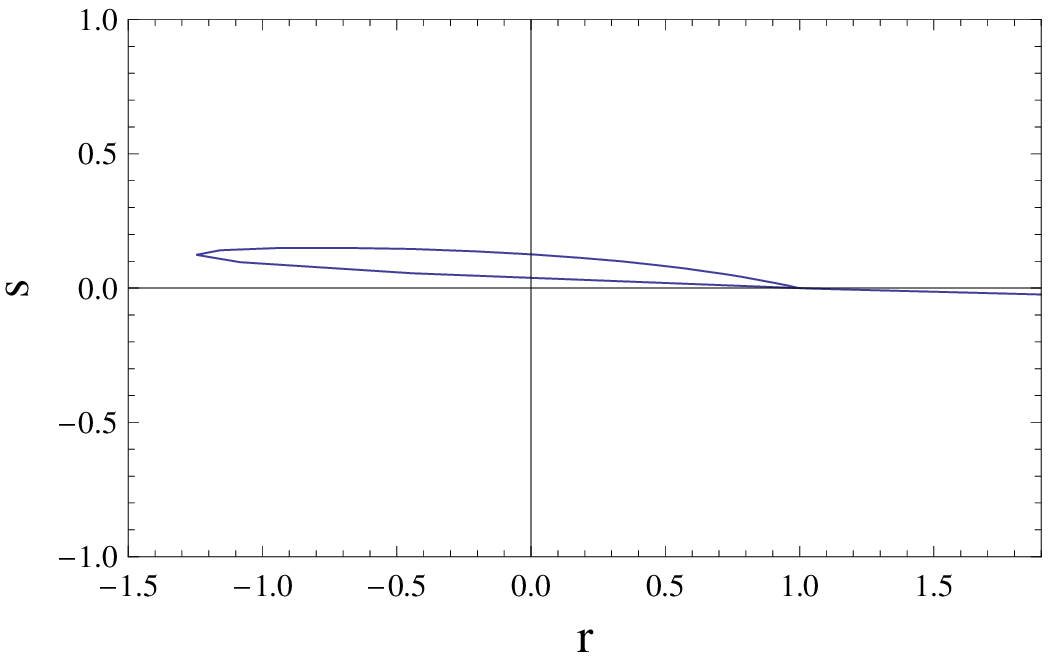}~~~~

\vspace{2mm}
~~~~~~~~~~~~~~~~~~Fig.12~~~~~~~~~~~~~~~~~~~\\
\vspace{6mm} Fig. 12 shows the variations of $r$ against $s$, for
$\alpha =0.02, \beta =10, \mu =10, \nu =75, m =10, n =.2, d =15$
in Emergent Scenario. \vspace{6mm}
\end{figure}

The relation between $r$ and $s$ has been shown in Fig.12. From
Fig.12, we see that $s$ is negative when $r\geq 1$. The curve
shows that the Universe starts from Einstein static era and goes
to the $\Lambda CDM$ model ($r=1,~s=0$).\\\\\\\\\\\\

\section{\normalsize\bf{Logamediate Scenario}}

Now we consider Logamediate scenario, where the scale factors
$a(t)$ and $b(t)$ are consider as the power of cosmic time $t$ are
given by [24]
\begin{equation}
a(t)=e^{A(\ln t)^{\lambda_{1}}}  \quad and  \quad  b(t)=e^{B(\ln
t)^{\lambda_{2}}}
\end{equation}
where $A$, $B$, $m$ and $n$ are positive constants. So the field
equations (5), (6) and (7) become
\begin{equation}
\frac{3A^{2}\lambda_{1}^{2}(\ln t)^{2\lambda_{1}-2}}{t^{2}}+3k
e^{-2A(\ln t)^{\lambda_{1}}}=\frac{B d \lambda_{2}(\ln
t)^{\lambda_{2}-2}}{8t^{2}}\left(4(\lambda_{2}-1)+B
\lambda_{2}(d+1)(\ln t)^{\lambda_{2}}-4\ln t\right)+\rho
\end{equation}

\begin{equation}
\frac{A\lambda_{1}(\ln t)^{\lambda_{1}-2}(2(\lambda_{1}-1)+3A
\lambda_{1}(\ln t)^{\lambda_{1}})}{t^{2}}+k e^{-2A(\ln
t)^{\lambda_{1}}}=\frac{A \lambda_{1}(\ln t)^{\lambda_{1}-1}(4+Bd
\lambda_{2}(\ln
t)^{\lambda_{2}-1})}{2t^{2}}+\frac{B^{2}\lambda_{2}^{2}d(d-1)(\ln
t)^{2\lambda_{2}-2}}{8t^{2}}-p
\end{equation}

\begin{equation}
\frac{B\lambda_{2}(\ln t)^{\lambda_{2}-2}(6A\lambda_{1} (\ln
t)^{\lambda_{1}})+Bd\lambda_{2}(\ln t)^{\lambda_{2}}-2\ln
t+2\lambda_{2}-2}{t^{2}}+p=0
\end{equation}

We now consider K-essence field, Tachyonic field,  normal scalar
field and DBI essence field. For these four cases we analyze the
behavior of the Logamediate Universe in extra dimension and
finally we analyze the behavior of the state finder parameters $r$ and $s$.\\

$\bullet$ \textbf{K-essence Field:}\\

The energy density and pressure due to K-essence field $\phi$ are
given by the equations (12) and (13). Using equations (38)-(40),
we can find the expressions for $V(\phi)$ and $\phi$ as

\begin{figure}
\includegraphics[scale=0.4]{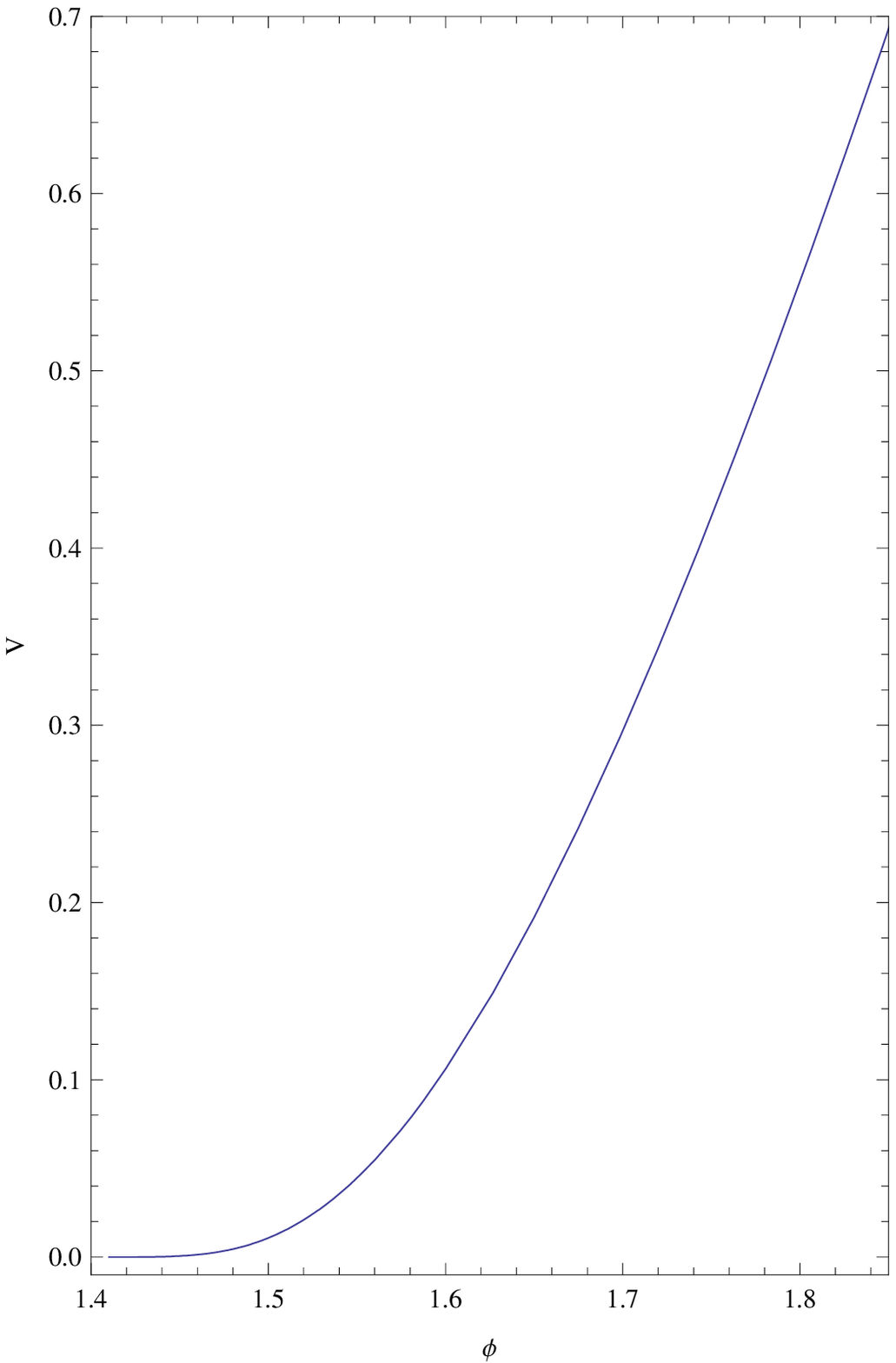}~~~~
\includegraphics[scale=0.7]{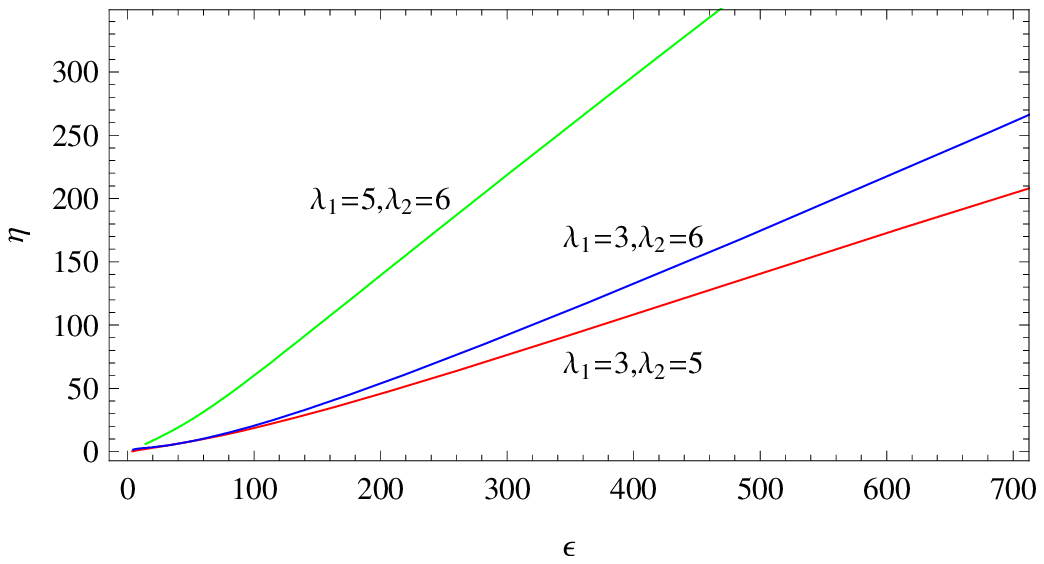}
\vspace{2mm}
~~~~~Fig.13~~~~~~~~~~~~~~~~~~~~~~~~~~~~~~~~~~~~~~~~~~~~~~~~~~~~~Fig.14~~~~~~~~~~~~~~~~~\\
\vspace{6mm} Fig. 13 shows the variations of $V$ against $\phi$,
for $A=1, B=2, k=1, \lambda_{1}=2, \lambda_{2}=3, d =5 $ and Fig.
14 shows the variation of the slow roll parameters $\epsilon$
against $\eta$ for $\lambda_{1}=3, \lambda_{2}=6, d =5 $ $k= -1,
1, 0$ respectively in the case of K-essence Scalar field for
Logamediate Scenario.
\vspace{6mm}
\end{figure}

\begin{eqnarray*}
V(\phi)=\frac{e^{-2A(\ln t)^{\lambda_{1}}}\left[B^{2}d
\lambda_{2}^{2}(2d-1)(\ln t)^{2\lambda_{2}}e^{2A(\ln
t)^{\lambda_{1}}}+2Bd\lambda_{2}(\ln t)^{\lambda_{2}}e^{2A(\ln
t)^{\lambda_{1}}} \left(3A\lambda_{1}(\ln t)^{\lambda_{1}}-\ln
t+\lambda_{2}-1\right)\right.}{8t^{2}(\ln
t)^{2}\left(-16kt^{2}(\ln t)^{2}-8A\lambda_{1} e^{2A(\ln
t)^{\lambda_{1}}}(\lambda_{1}-1-\ln t)(\ln
t)^{\lambda_{1}}-24A^{2}\lambda_{1}^{2}(\ln
t)^{2\lambda_{1}}e^{2A(\ln t)^{\lambda_{1}}}\right.}
\end{eqnarray*}
\begin{equation}
\frac{\left.-24\left(kt^{2}(\ln t)^{2}+A\lambda_{1} e^{2A(\ln
t)^{\lambda_{1}}}(\lambda_{1}-1-\ln t)(\ln
t)^{\lambda_{1}}+2A^{2}\lambda_{1}^{2}(\ln
t)^{2\lambda_{1}}e^{2A(\ln
t)^{\lambda_{1}}}\right)\right]^{2}}{\left.+B^{2}d^{2}\lambda_{2}^{2}(\ln
t)^{2\lambda_{2}}e^{2A(\ln t)^{\lambda_{1}}}+2Bd\lambda_{2}(\ln
t)^{\lambda_{2}}e^{2A(\ln t)^{\lambda_{1}}}\left(A\lambda_{1}(\ln
t)^{\lambda_{1}}-\ln t+\lambda_{2}-1\right)\right)}
\end{equation}
and
\begin{eqnarray*}
\phi=\int{\left[\frac{-2B^{2}d^{2}\lambda_{2}^{2}(\ln
t)^{2\lambda_{2}}e^{2A(\ln t)^{\lambda_{1}}}-4Bd\lambda_{2}(\ln
t)^{\lambda_{2}}e^{2A(\ln t)^{\lambda_{1}}}\left(A\lambda_{1}(\ln
t)^{\lambda_{1}}-\ln t+\lambda_{2}-1\right)+16\left(2kt^{2}(\ln
t)^{2}\right.}{-B^{2}d\lambda_{2}^{2}(2d-1)(\ln
t)^{2\lambda_{2}}e^{2A(\ln t)^{\lambda_{1}}}-2Bd\lambda_{2}(\ln
t)^{\lambda_{2}}e^{2A(\ln t)^{\lambda_{1}}}\left(3A\lambda_{1}(\ln
t)^{\lambda_{1}}-\ln t+\lambda_{2}-1\right)}\right.}
\end{eqnarray*}
\begin{equation}
\left.\frac{\left.+A\lambda_{1}e^{2A(\ln
t)^{\lambda_{1}}}(\lambda_{1}-1-\ln t)(\ln
t)^{\lambda_{1}}+3A^{2}\lambda_{1}^{2}(\ln
t)^{2\lambda_{1}}e^{2A(\ln
t)^{\lambda_{1}}}\right)}{+24\left(kt^{2}(\ln
t)^{2}+A\lambda_{1}e^{2A(\ln t)^{\lambda_{1}}}(\lambda_{1}-1-\ln
t)(\ln t)^{\lambda_{1}}+2A^{2}\lambda_{1}^{2}(\ln
t)^{2\lambda_{1}}e^{2A(\ln
t)^{\lambda_{1}}}\right)}\right]^{\frac{1}{2}}~dt
\end{equation}

$\bullet$ \textbf{Tachyonic  field:}\\

The energy density $\rho$ and pressure $p$ due to the Tachyonic
field $\phi$ are given by the equations (16) and (17). Using
equations (38)-(40), we can find the expressions for $V(\phi)$ and
$\phi$ as

\begin{eqnarray*}
V(\phi)=\sqrt{\frac{\left(24kt^{2}(\ln t)^{2}e^{-2A(\ln
t)^{\lambda_{1}}}+24A^{2}\lambda_{1}^{2}(\ln
t)^{2\lambda_{1}}-4Bd\lambda_{2}(\lambda_{2}-1)(\ln
t)^{\lambda_{2}}-B^{2}d(d+1)\lambda_{2}^{2}(\ln
t)^{2\lambda_{2}}+4Bd\lambda_{2}(\ln
t)^{\lambda_{2}+1}\right)}{64t^{4}(\ln t)^{4}}}
\end{eqnarray*}
\begin{eqnarray*}
\times\sqrt{\left[8kt^{2}(\ln t)^{2}e^{-2A(\ln
t)^{\lambda_{1}}}+16A\lambda_{1}(\lambda_{1}-1)(\ln
t)^{\lambda_{1}}+24A^{2}\lambda_{1}^{2}(\ln
t)^{2\lambda_{1}}-16A\lambda_{1}(\ln
t)^{\lambda_{1}+1}-B^{2}d(d-1)\lambda_{2}^{2}(\ln
t)^{2\lambda_{2}}\right.}
\end{eqnarray*}
\begin{equation}
\overline{\left.-4ABd\lambda_{1}\lambda_{2}(\ln
t)^{\lambda_{1}+\lambda_{2}}\right]}
\end{equation}

and
\begin{eqnarray*}
\phi=\int{\sqrt{\left[1-\frac{8kt^{2}(\ln t)^{2}e^{-2A(\ln
t)^{\lambda_{1}}}+16A\lambda_{1}(\lambda_{1}-1)(\ln
t)^{\lambda_{1}}+24A^{2}\lambda_{1}^{2}(\ln
t)^{2\lambda_{1}}-16A\lambda_{1}(\ln
t)^{\lambda_{1}+1}-B^{2}d(d-1)\lambda_{2}^{2}(\ln
t)^{2\lambda_{2}}}{24kt^{2}(\ln t)^{2}e^{-2A(\ln
t)^{\lambda_{1}}}+24A^{2}\lambda_{1}^{2}(\ln
t)^{2\lambda_{1}}-4Bd\lambda_{2}(\lambda_{2}-1)(\ln
t)^{\lambda_{2}}-B^{2}d(d+1)\lambda_{2}^{2}(\ln
t)^{2\lambda_{2}}}\right.}}
\end{eqnarray*}

\begin{equation}
\overline{{\left.\frac{-4ABd\lambda_{1}\lambda_{2}(\ln
t)^{\lambda_{1}+\lambda_{2}}}{+4Bd\lambda_{2}(\ln
t)^{\lambda_{2}+1}}\right]}}~dt
\end{equation}

\begin{figure}
\includegraphics[scale=0.5]{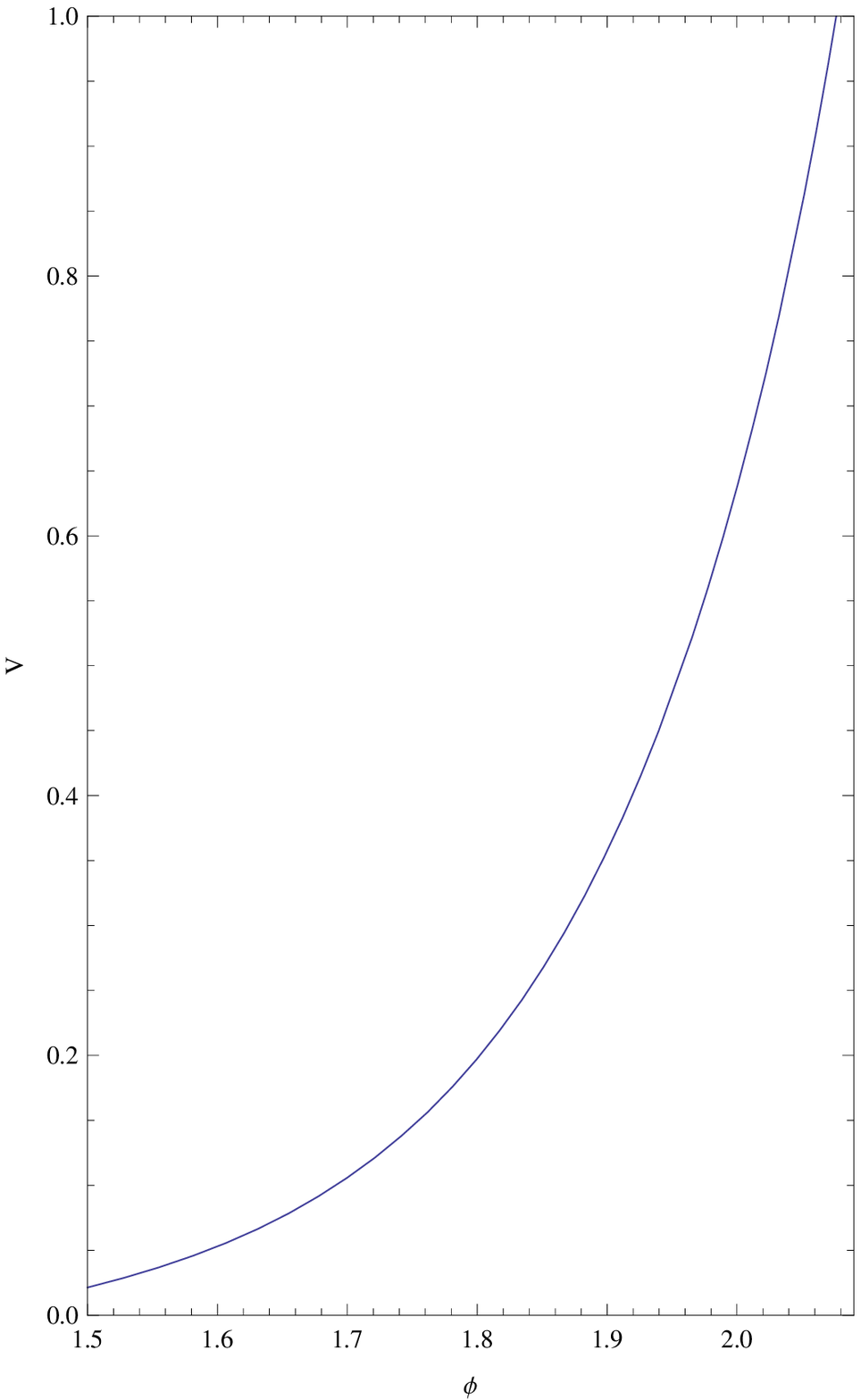}~~~~
\includegraphics[scale=0.65]{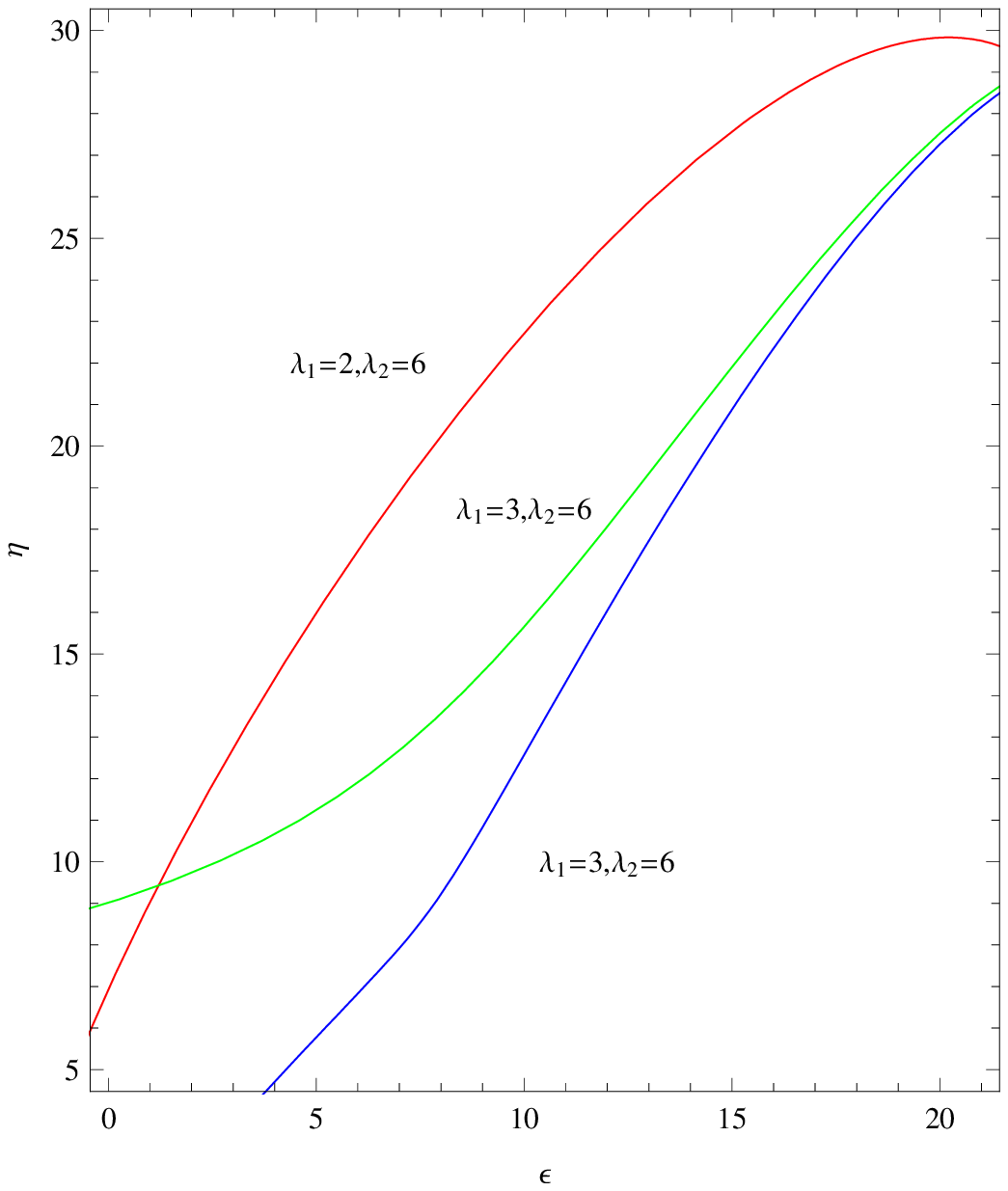}
\vspace{2mm}
~~~~Fig.15~~~~~~~~~~~~~~~~~~~~~~~~~~~~~~~~~~~~~~~~~~~~~~~~Fig.16~~~~~~~~~~~~\\
\vspace{6mm} Fig. 15 shows the variations of $V$ against $\phi$,
for $A=1, B=2, k=1, \lambda_{1}=5, \lambda_{2}=4, d=15$ and Fig.
16 shows the variation of the slow roll parameters $\epsilon$
against $\eta$ for $\lambda_{1}=3, \lambda_{2}=7, d=5$ $k= -1, 1,
0$ respectively in the case of Tachyonic Scalar field for
Logamediate Scenario.
\vspace{6mm}
\end{figure}

$\bullet$ \textbf{Normal Scalar field:}\\

The energy density $\rho$ and pressure $p$ due to the Normal
Scalar field $\phi$ are given by the equations (20) and (21).
Using equations (38)-(40), we can find the expressions for
$V(\phi)$ and $\phi$ as

\begin{eqnarray*}
V(\phi)=\frac{1}{8t^{2}(\ln t)^{2}}\left[16kt^{2}(\ln
t)^{2}e^{-2A(\ln
t)^{\lambda_{1}}}+8A\lambda_{1}(\lambda_{1}-1)(\ln
t)^{\lambda_{1}}+24A^{2}\lambda_{1}^{2}(\ln
t)^{2\lambda_{1}}-8A\lambda_{1}(\ln t)^{\lambda_{1}+1}\right.
\end{eqnarray*}
\begin{equation}
\left.-2Bd\lambda_{2}(\lambda_{2}-1)(\ln
t)^{\lambda_{2}}-B^{2}d^{2}\lambda_{2}^{2}(\ln
t)^{2\lambda_{2}}+2Bd\lambda_{2}(\ln
t)^{\lambda_{2}+1}-2ABd\lambda_{1}\lambda_{2}(\ln
t)^{\lambda_{1}+\lambda_{2}}\right]
\end{equation}
and
\begin{eqnarray*}
\phi=\int{\sqrt{\frac{1}{4t^{2}(\ln t)^{2}}\left[8kt^{2}(\ln
t)^{2}e^{-2A(\ln t)^{\lambda_{1}}}
+8A\lambda_{1}(\lambda_{1}-1)(\ln t)^{\lambda_{1}}
+8A\lambda_{1}(\ln t)^{\lambda_{1}+1}
-2Bd\lambda_{2}(\lambda_{2}-1)(\ln t)^{\lambda_{2}}\right.}}
\end{eqnarray*}
\begin{equation}
\overline{\left.-B^{2}d^{2}\lambda_{2}^{2}(\ln
t)^{2\lambda_{2}}+2Bd\lambda_{2}(\ln
t)^{\lambda_{2}+1}-2ABd\lambda_{1}\lambda_{2}(\ln
t)^{\lambda_{1}+\lambda_{2}}\right]}~dt
\end{equation}

\begin{figure}
\includegraphics[scale=0.5]{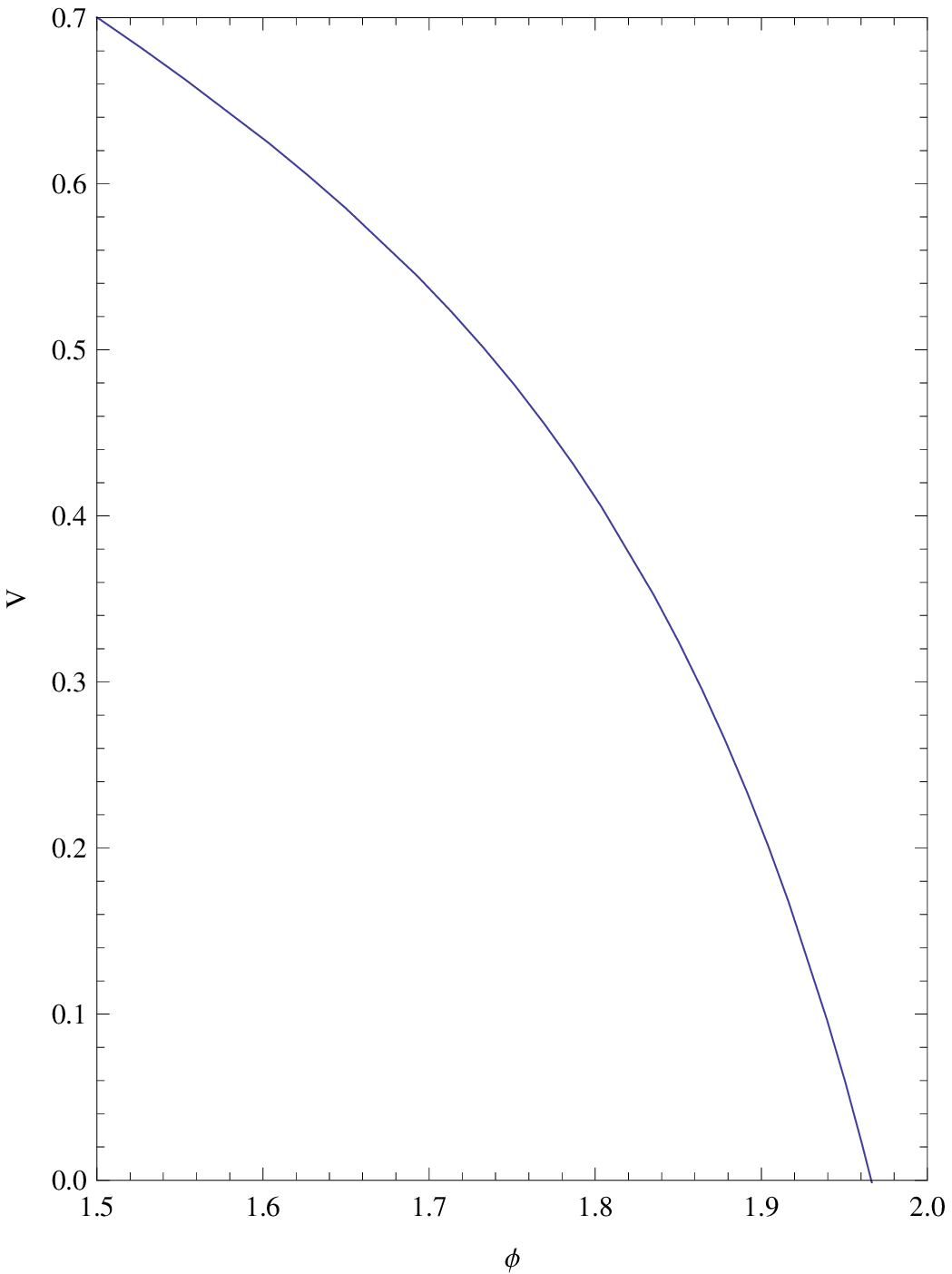}~~~~
\includegraphics[scale=0.9]{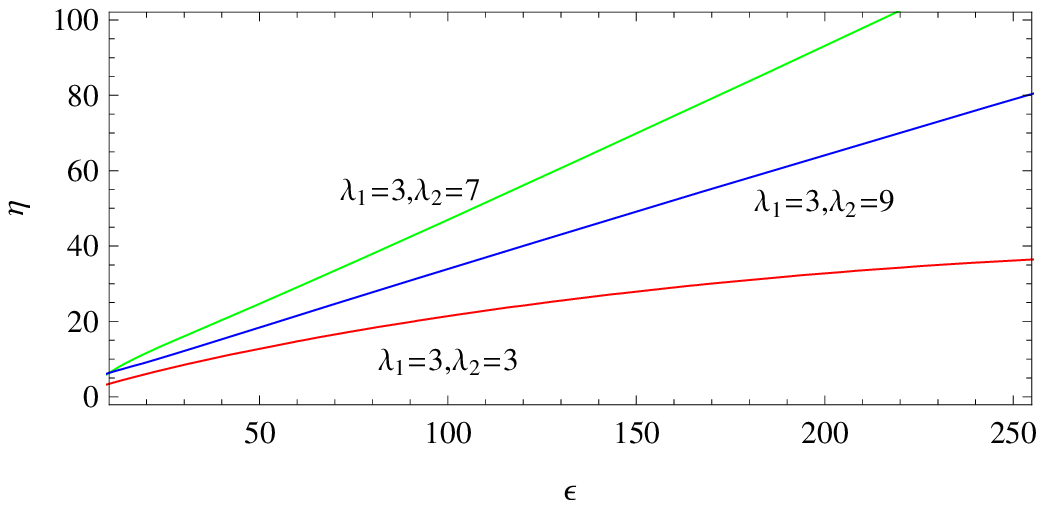}
\vspace{2mm}
Fig.17~~~~~~~~~~~~~~~~~~~~~~~~~~~~~~~~~~~~~~~~~~~~~~~~~~~~~~~~~~~~~~~~~~~~~~Fig.18~~~~~~~~~~\\
\vspace{6mm} Fig. 17 shows the variations of $V$ against $\phi$,
for $A=0.01, B=0.2, k=1, \lambda_{1}=5, \lambda_{2}=4, d=15$ and
Fig. 18 shows the variation of the slow roll parameters $\epsilon$
against $\eta$ for $\lambda_{1}=2, \lambda_{2}=3, d=5$ $k= -1, 1,
0$ respectively in the case of Normal Scalar field for Logamediate
Scenario.
\vspace{6mm}
\end{figure}

$\bullet$ \textbf{DBI-essence:}\\

The energy density $\rho$ and pressure $p$ due to the DBI-essence
field $\phi$ are given by the equations (24) and (25).\\

{\bf Case I:}  $\gamma$ = constant.\\
 Using equations (38)-(40), we can find the expressions for $V(\phi)$,
$T(\phi)$ and $\phi$ as

\begin{equation}
V(\phi)=\delta \phi_{0}^{2}e^{(2AF(\ln t)^{\lambda_{1}}+2BG(\ln
t)^{\lambda_{2}})}
\end{equation}

\begin{equation}
T(\phi)=\sigma\phi_{0}^{2}e^{(2AF(\ln t)^{\lambda_{1}}+2BG(\ln
t)^{\lambda_{2}})}
\end{equation}
and
\begin{equation}
\phi=\int{\phi_{0}e^{(AF(\ln t)^{\lambda_{1}}+BG(\ln
t)^{\lambda_{2}})}}~dt
\end{equation}

where$E=\frac{1}{\gamma}[2(\delta-\sigma)+2(\sigma-1)\gamma^{3}]$,
$\gamma=\sqrt{\frac{\sigma}{\sigma-1}}$, $F=-\frac{3}{E}$,
$G=-\frac{d}{E}$ and $\phi_{0}$ is an integrating constants. From
above, we see that $V(\phi)$ and $T(\phi)$ are both exponentially
decreasing with DBI scalar field
$\phi$.\\

\begin{figure}
\includegraphics[scale=0.6]{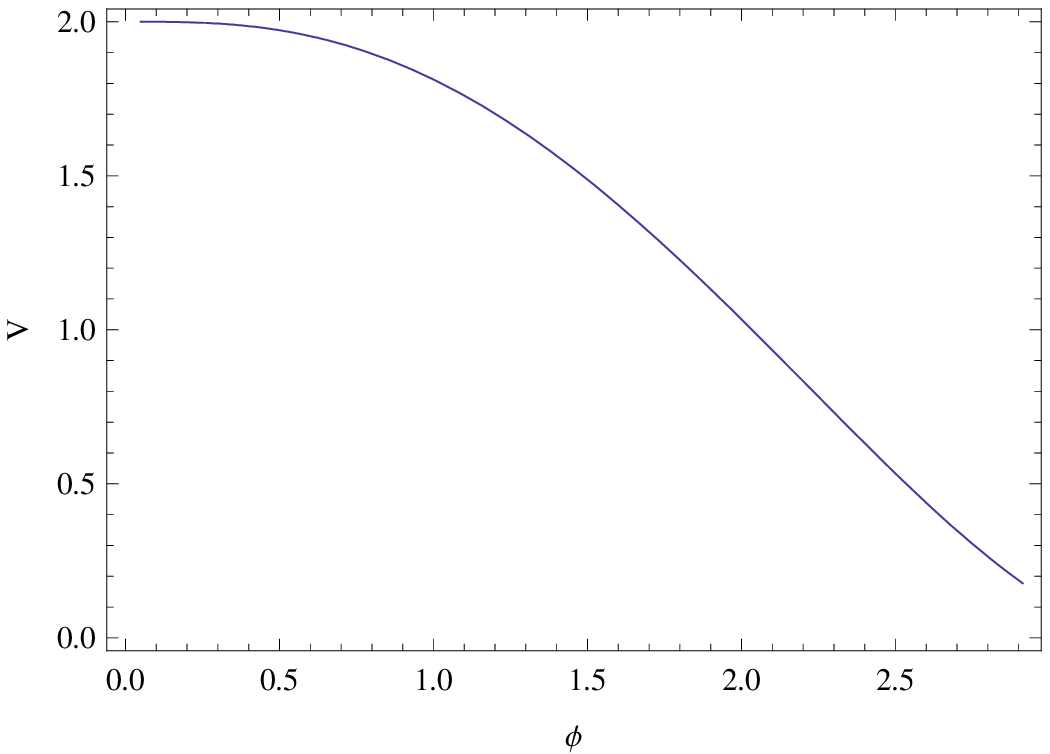}~~~~~~~~~~~~~~~
\includegraphics[scale=0.5]{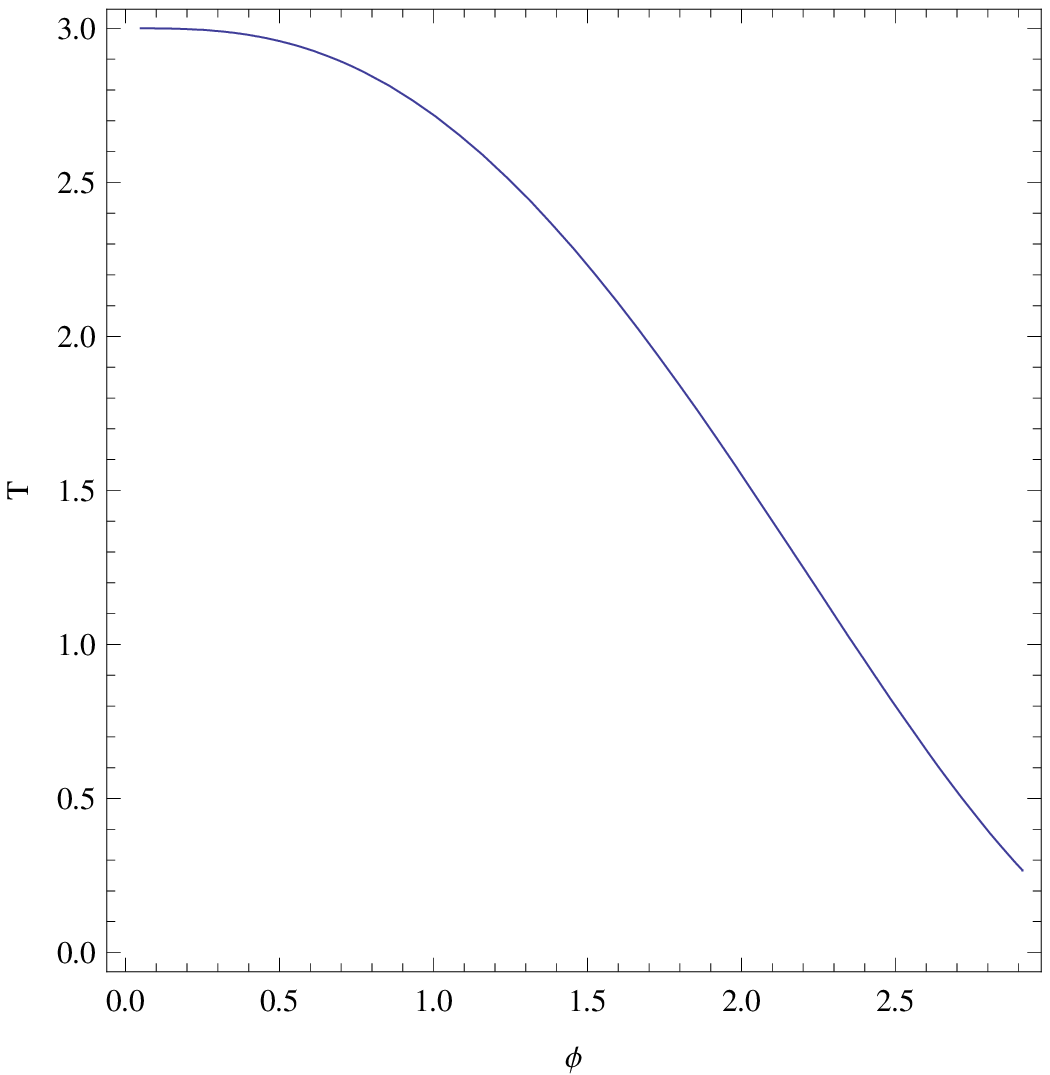}\\
\vspace{2mm}
~~~~~~~~~~~~~~Fig.19~~~~~~~~~~~~~~~~~~~~~~~~~~~~~~~~~~~~~~~~~~~~~~~~~~~~~~~~~~~~~~Fig.20~~~~\\
\vspace{4mm}
\includegraphics[scale=.8]{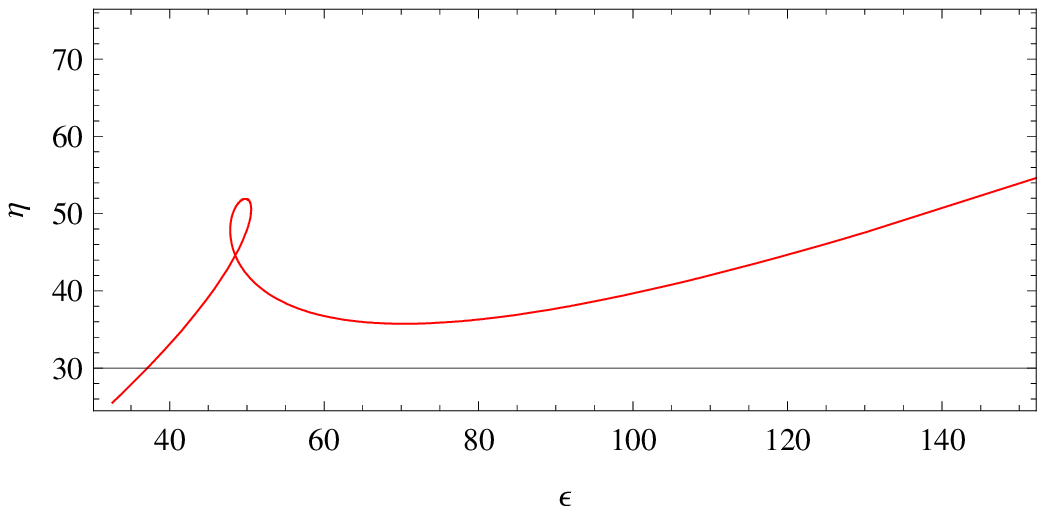}\\
\vspace{2mm}
~~~~~~~~~~~~~~~~~~~~~~~~~~~~~~~~~~~~~~~~~~Fig.21~~~~~~~~~~~~~~~~~~~~~~~~~~~~~~~~\\
\vspace{6mm} Fig. 19 shows the variations of $V$ against $\phi$
and Fig. 20 shows the variations of $T$ against $\phi$, for
$A=0.3, B=0.2,\lambda_{1}=4, \lambda_{2}=3, d=15, \sigma=3,
\delta=2,\phi_{0}=2$ and Fig. 21 shows the variation of the slow
roll parameters $\epsilon$ against $\eta$ for
$A=0.01,B=0.02,\lambda_{1}=3, \lambda_{2}=8,\sigma=8,\delta=2,
\phi_{0}=1,d=20$ in the $1st$ case of DBI-essence Scalar field for
Logamediate Scenario.
 \vspace{6mm}
\end{figure}

{\bf Case II:}  $\gamma \neq$  constant.\\

Using equations (38)-(40), we can find the expressions for
$V(\phi)$ and $\phi$ as

\begin{equation}
V(\phi)=\ln{[\frac{C_{0}}{e^{(3A(\ln t)^{\lambda_{1}}+dB(\ln
t)^{\lambda_{2}})}}]}\sqrt{1-\frac{1}{\ln{[\frac{C_{0}}{e^{(3A(\ln
t)^{\lambda_{1}}+dB(\ln t)^{\lambda_{2}})}}]}}}
\end{equation}
and
\begin{equation}
\phi=\int{\left(1-\frac{1}{\ln{[\frac{C_{0}}{e^{(3A(\ln
t)^{\lambda_{1}}+dB(\ln
t)^{\lambda_{2}})}}]}}\right)^{\frac{1}{4}}}~dt
\end{equation}
Where $C_{0}$ is an integrating constant.\\\\

\begin{figure}
\includegraphics[scale=0.45]{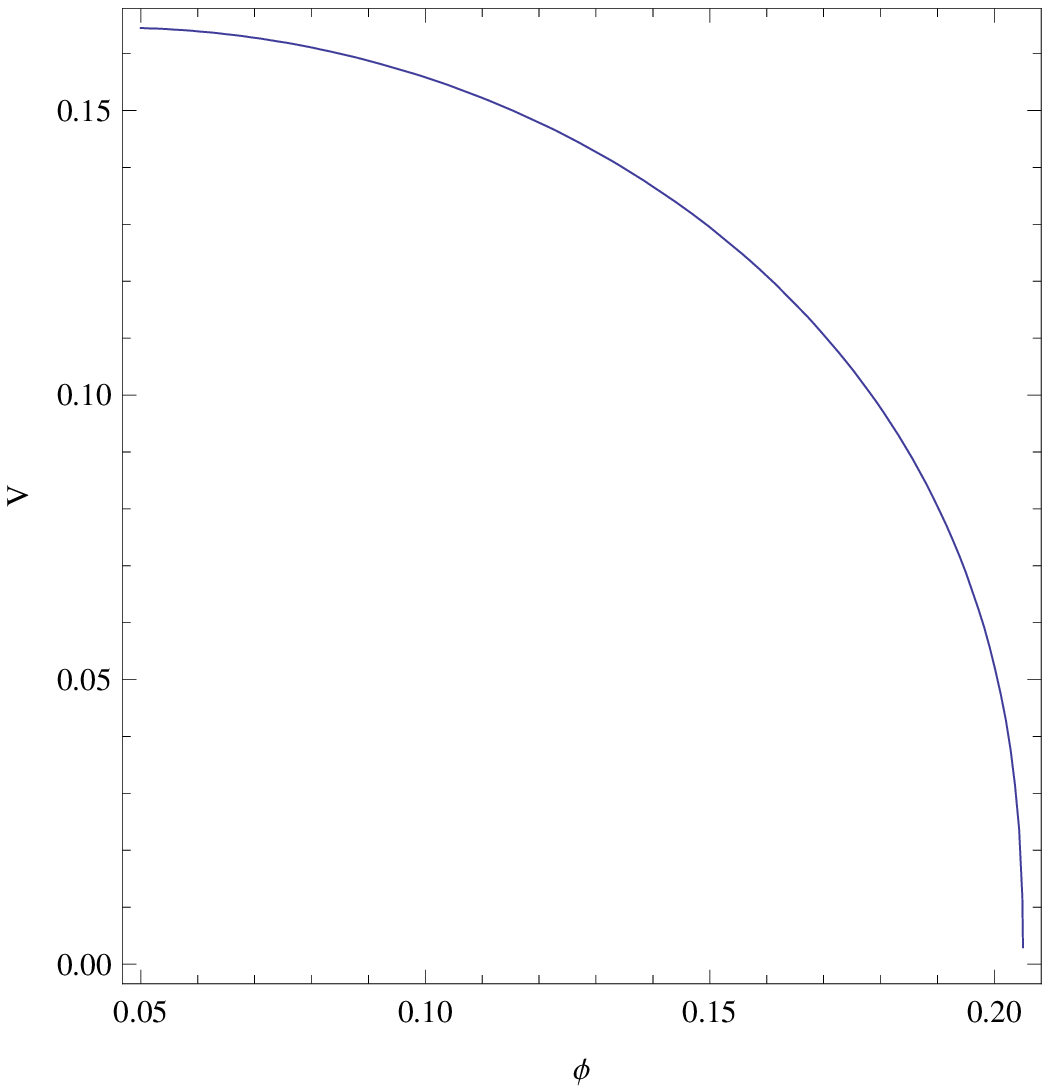}~~~~
\includegraphics[scale=.5]{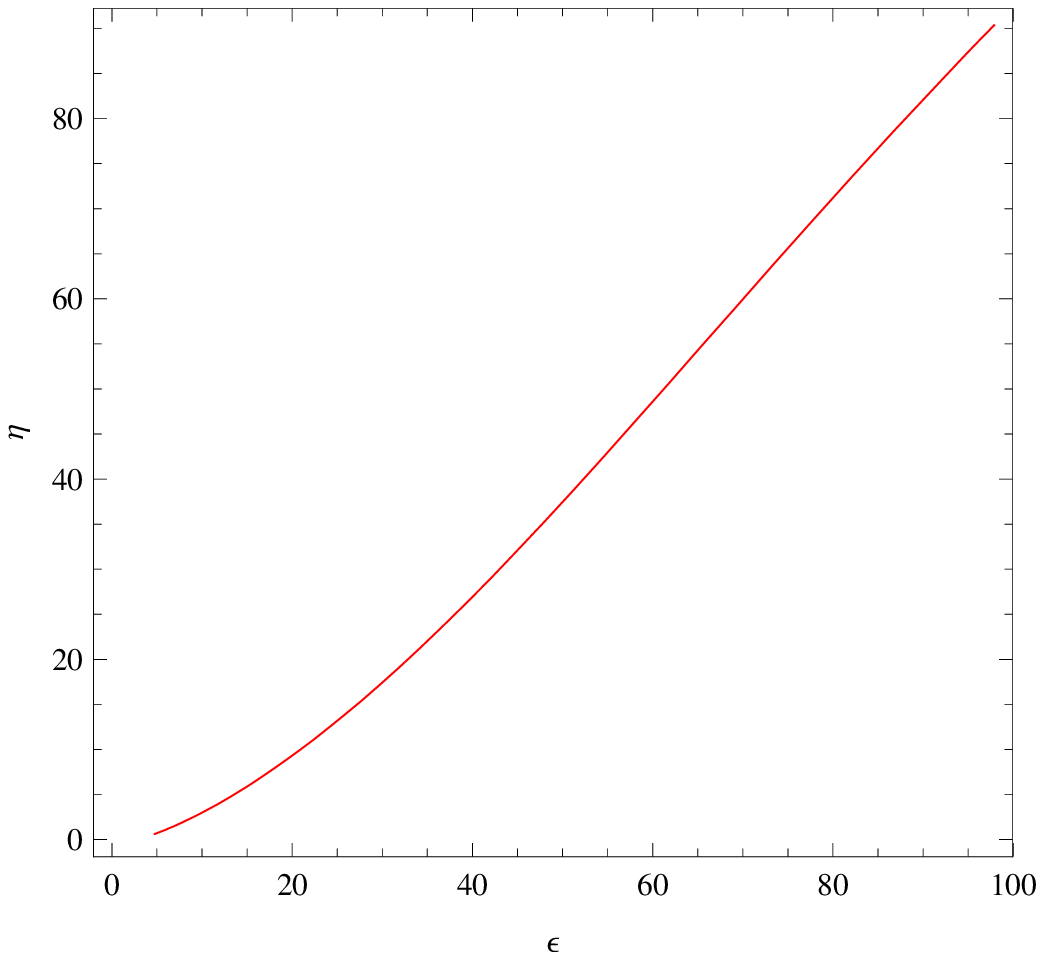}\\
\vspace{2mm}
~~~~~~Fig.22~~~~~~~~~~~~~~~~~~~~~~~~~~~~~~~~~~~~~~~~~~~~~~~~~~Fig.23~~~~~~\\
\vspace{6mm} Fig. 22 shows the variations of $V$ against $\phi$,
for $A=0.3, B=0.2,C_{0}=3,\lambda_{1}=2, \lambda_{2}=3, d=5$  and
Fig. 23 shows the variation of the slow roll parameters $\epsilon$
against $\eta$ for $A=2,B=3,C_{0}=1,\lambda_{1}=3, \lambda_{2}=4,
d=5$ in the $2nd$ case of DBI-essence Scalar field for Logamediate
Scenario.
 \vspace{6mm}
\end{figure}

$\bullet$ \textbf{Statefinder parameters:}\\

The geometrical parameters\{$r,s$\} for higher dimensional
anisotropic cosmology in Logamediate scenario can be constructed
from the scale factors $a(t)$ and $b(t)$ as

\begin{eqnarray*}
r=1+\frac{(3(d+3)(3A\lambda_{1}(-\ln t+\lambda_{1}-1)(\ln
t)^{\lambda_{1}}+B d \lambda_{2}(-\ln t+\lambda_{2}-1)(\ln
t)^{\lambda_{2}}))}{(3A\lambda_{1}(\ln t)^{\lambda_{1}}+B d
\lambda_{2}(\ln t)^{\lambda_{2}})^{2}}
\end{eqnarray*}
\begin{equation}
+\frac{(d+3)^{2}(3A\lambda_{1}(\ln
t)^{\lambda_{1}}(2+\lambda_{1}(\lambda_{1}-3)+\ln t(2\ln
t-3\lambda_{1}+3))+Bd\lambda_{2}(\ln
t)^{\lambda_{2}}(2+\lambda_{1}(\lambda_{1}-3)+\ln t(2\ln
t-3\lambda_{1}+3)))}{(3A\lambda_{1}(\ln
t)^{\lambda_{1}}+Bd\lambda_{2}(\ln t)^{\lambda_{2}})^{3}}
\end{equation}
\begin{eqnarray*}
s=\frac{(d+3)3(3A\lambda_{1}(\ln t)^{\lambda_{1}}+B d
\lambda_{2}(\ln t)^{\lambda_{2}})(3A\lambda_{1}(-\ln
t+\lambda_{1}-1)(\ln t)^{\lambda_{1}}+B d \lambda_{2}(-\ln
t+\lambda_{2}-1)(\ln t)^{\lambda_{2}})}{\left[3(3A\lambda_{1}(\ln
t)^{\lambda_{1}}+B d \lambda_{2}(\ln t)^{\lambda_{2}})^{3}\right.}
\end{eqnarray*}
\begin{equation}
\frac{\left.+(d+3)(3A\lambda_{1}(\ln
t)^{\lambda_{1}}(2+\lambda_{1}(\lambda_{1}-3)+\ln t(2\ln
t-3\lambda_{1}+3))+Bd \lambda_{2}(\ln
t)^{\lambda_{2}}(2+\lambda_{2}(\lambda_{2}-3)+\ln t(2\ln
t-3\lambda_{2}+3)))\right]}{\left.(-\frac{3}{2}+\frac{((d+3)(-3A\lambda_{1}(-\ln
t+\lambda_{1}-1)(\ln t)^{\lambda_{1}}-B d \lambda_{2}(-\ln
t+\lambda_{2}-1)(\ln t)^{\lambda_{2}}))}{(3A\lambda_{1}(\ln
t)^{\lambda_{1}}+B d \lambda_{2}(\ln
t)^{\lambda_{2}})^{2}})\right]}
\end{equation}

\begin{figure}
\includegraphics[scale=0.9]{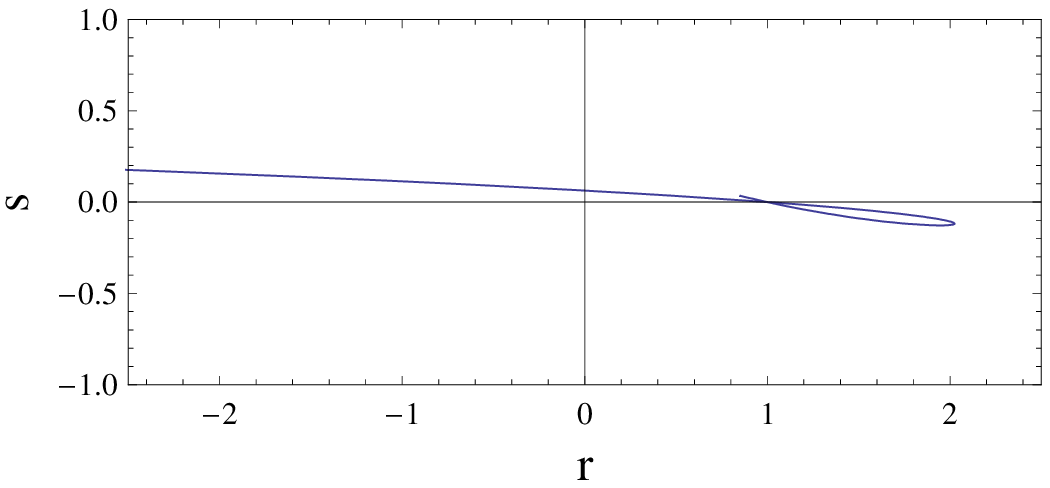}~~~~

\vspace{2mm}
~~~~~~~~~~~~~~~~~~Fig.24~~~~~~~~~~~~~~~~~~~\\
\vspace{6mm} Fig. 24 shows the variations of $r$ against $s$, for
$A=1.5, B=3.3, \lambda_{1}=2, \lambda_{2}=2, d =3 $ in Logamediate
Scenario.
\vspace{6mm}
\end{figure}

\section{\normalsize\bf{Intermediate  Scenario}}

Finally we consider Intermediate scenario, where the scale factors
$a(t)$ and $b(t)$ are consider as the power of cosmic time $t$ are
given by [24]
\begin{equation}
a(t)=e^{A t^{f_{1}}}  \quad and  \quad  b(t)=e^{B t^{f_{2}}}
\end{equation}
where $A$, $B$, $m$ and $n$ are positive constants. So the field
equations (3), (4) and (5) become
\begin{equation}
\frac{1}{8t^{2}}e^{-2At^{f_{1}}}(24kt^{2}+e^{2At^{f_{1}}}
(24A^{2}f_{1}^{2}t^{2f_{1}}-Bd
f_{2}t^{f_{2}}(-4+f_{2}(4+B(d+1)t^{f_{2}}))))-\rho=0
\end{equation}

\begin{equation}
\frac{1}{8t^{2}}e^{-2At^{f_{1}}}(-8kt^{2}+e^{2At^{f_{1}}}(-24A^{2}f_{1}^{2}t^{2f_{1}}+
B^{2}(d-1)d
f_{2}^{2}t^{2f_{2}}-4Af_{1}t^{f_{1}}(4f_{1}-Bdf_{2}t^{f_{2}}-4)))-p=0
\end{equation}

\begin{equation}
Bf_{2}t^{f_{2}-2}(6Af_{1}t^{f_{1}}+n(Bdt^{f_{2}}+2)-2)+p=0
\end{equation}

We now consider K-essence field, Tachyonic field,  normal scalar
field and DBI essence field. For these four cases we analyze the
behavior of the Intermediate Universe in extra dimension and
finally we analyze the behavior of the state finder parameters $r$ and $s$.\\

$\bullet$ \textbf{K-essence Field:}\\

The energy density and pressure due to K-essence field $\phi$ are
given by the equations (12) and (13). Using equations (55)-(57),
we can find the expressions for $V(\phi)$ and $\phi$ as

\begin{equation}
V(\phi)
=\frac{-\left(e^{-2At^{f_{1}}}\left(24kt^{2}+e^{2At^{f_{1}}}\left(-B^{2}d(2d-1)f_{2}^{2}t^{2f_{2}}+
24Af_{1} t^{f_{1}}(2Af_{1}t^{f_{1}}+f_{1}-1)-2Bdf_{2}
t^{f_{2}}\left(3Af_{1}t^{f_{1}}+f_{2}
-1\right)\right)\right)^{2}\right)}{\left(8t^{2}\left(16kt^{2}+e^{2A
t^{f_{1}}}\left(-B^{2}d^{2}f_{2}^{2}t^{2f_{2}}-2 B d f_{2}
t^{f_{2}}(Af_{1}t^{f_{1}}+f_{2}-1)+8A f_{1} t^{f_{1}}(3A f_{1}
t^{f_{1}}+f_{1}-1)\right)\right)\right)}
\end{equation}
and
\begin{equation}
\phi=\int{\sqrt{\frac{2\left(16kt^{2}+e^{2A
t^{f_{1}}}\left(-B^{2}d^{2}f_{2}^{2}t^{2f_{2}}-2 B d f_{2}
t^{f_{2}}(Af_{1}t^{f_{1}}+f_{2}-1) + 8 A f_{1} t^{f_{1}}(3A f_{1}
t^{f_{1}}+f_{1}-1)\right)\right)}{\left(24kt^{2}+
e^{2At^{f_{1}}}\left(-B^{2}d(2d-1)f_{2}^{2}t^{2f_{2}}+ 24Af_{1}
t^{f_{1}}(2Af_{1}t^{f_{1}}+f_{1}-1)-2Bdf_{2}
t^{f_{2}}\left(3Af_{1}t^{f_{1}}+f_{2}
-1\right)\right)\right)}}}~dt
\end{equation}

\begin{figure}
\includegraphics[scale=0.5]{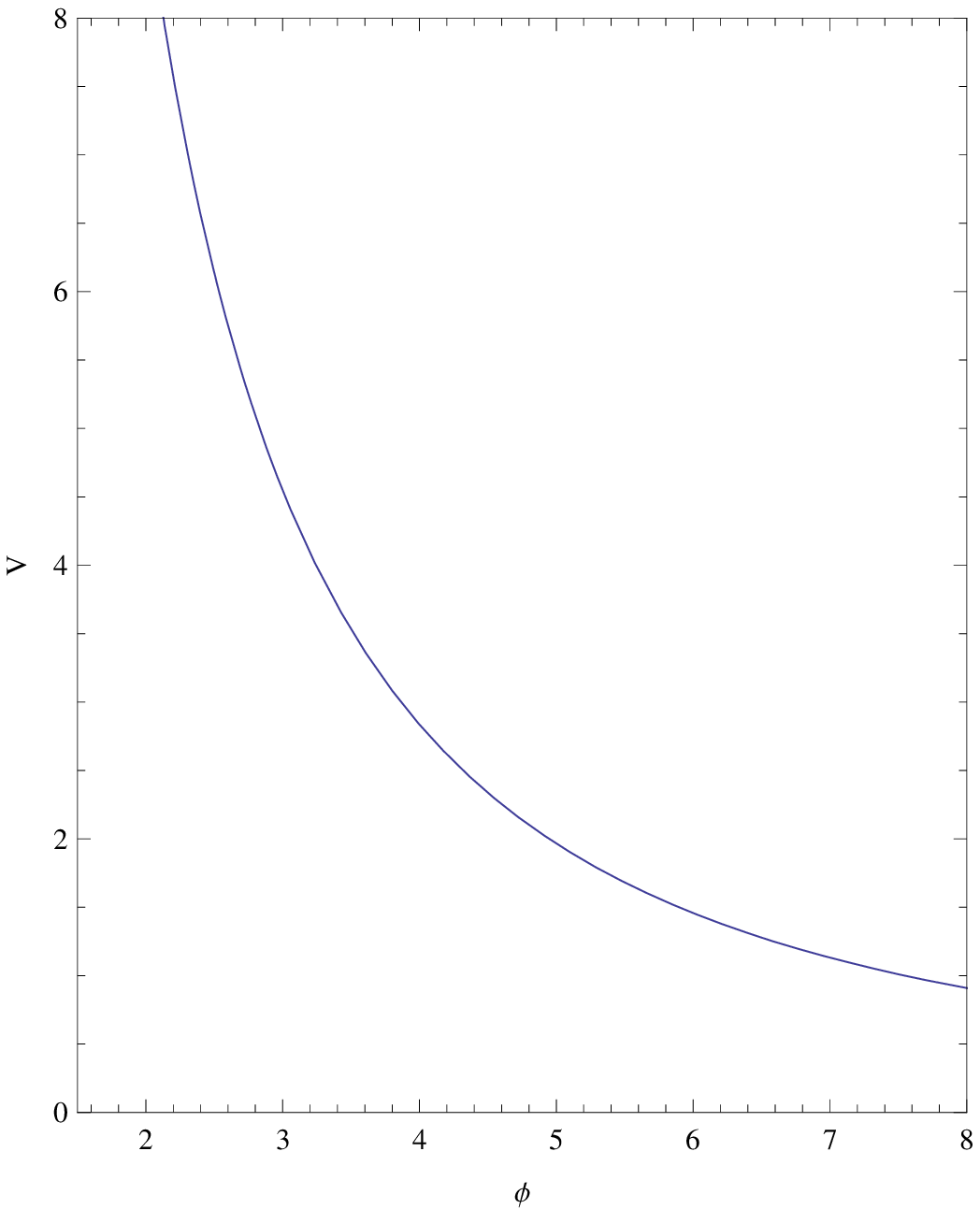}~~~~~
\includegraphics[scale=0.5]{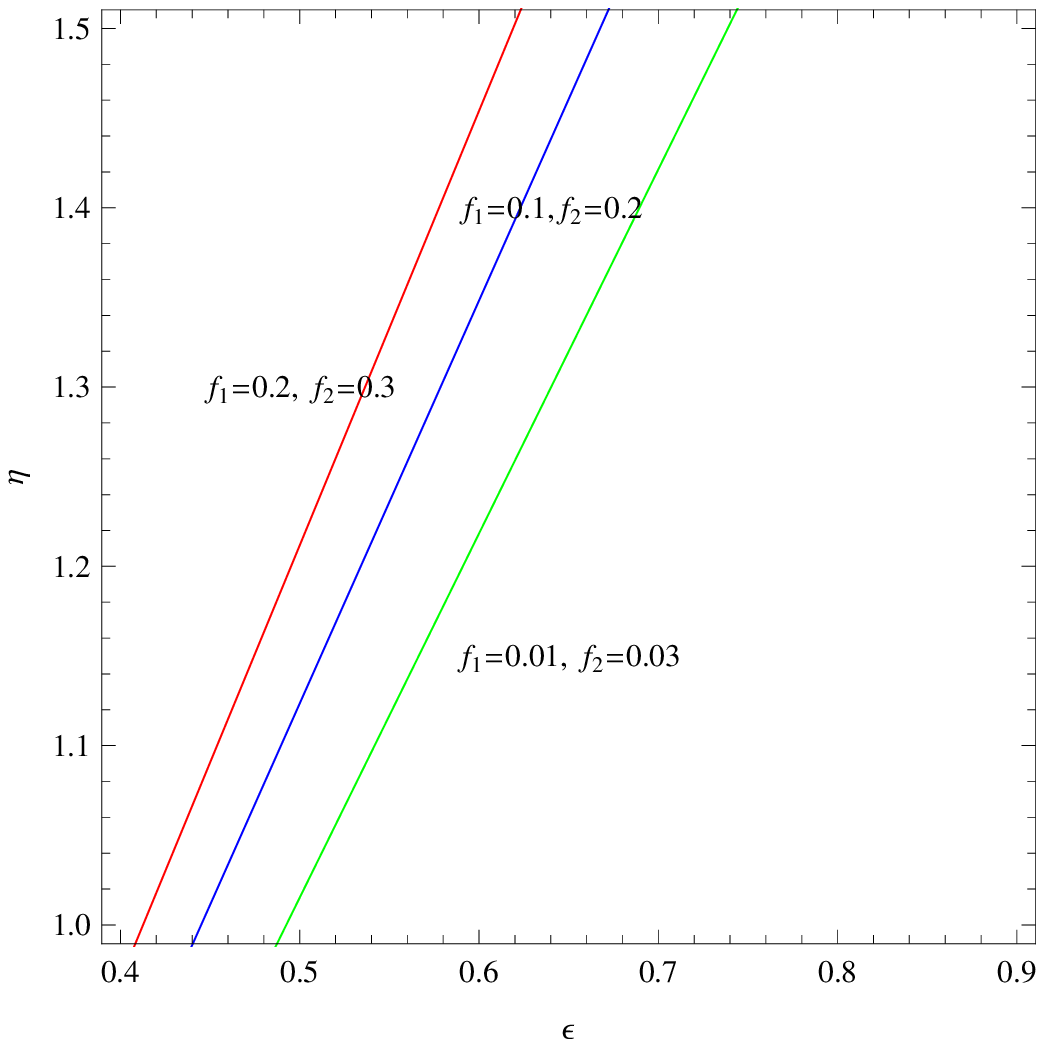}
\vspace{2mm}
~~~~~~~Fig.25~~~~~~~~~~~~~~~~~~~~~~~~~~~~~~~~~~~~~~~~~~~~~~~~~~~~Fig.26~~\\
\vspace{6mm} Fig. 25 shows the variations of $V$ against $\phi$,
for $A=20, B=25, k=1, f_{1}=0.2, f_{2}=0.19, d=5$ and Fig. 26
shows the variation of the slow roll parameters $\epsilon$ against
$\eta$ for $A =400, B=300, d=75$ and $k= -1, 1, 0$ respectively in
the case of K-essence Scalar field for Intermediate Scenario.
\vspace{6mm}
\end{figure}

$\bullet$ \textbf{Tachyonic  field:}\\

The energy density $\rho$ and pressure $p$ due to the Tachyonic
field $\phi$ are given by the equations (16) and (17). Using
equations (55)-(57), we can find the expressions for $V(\phi)$ and
$\phi$ as

\begin{eqnarray*}
V(\phi)=\frac{1}{8}\left[\frac{1}{t^{4}}
e^{-4At^{f_{1}}}\left(8kt^{2}+e^{2At^{f_{1}}}\left(-B^{2}d(d-1)
f_{2}^{2}t^{2f_{2}}-4ABdf_{1}f_{2}t^{f_{1}+f_{2}}+
8Af_{1}t^{f_{1}}\left(-2+f_{1}\left(3A
t^{f_{1}}+2\right)\right)\right)\right)\right.
\end{eqnarray*}
\begin{equation}
\left.\left(24kt^{2}+
e^{2At^{f_{1}}}\left(24A^{2}f_{1}^{2}t^{2f_{1}}-B d f_{2}
t^{f_{2}}\left(f_{2}-4\left(4+B(d+1)
t^{f_{2}}\right)\right)\right)\right)\right]^{\frac{1}{2}}
\end{equation}
and
\begin{equation}
\phi=\int{\sqrt{\frac{2\left(8kt^{2}+e^{2At^{f_{1}}}\left(-B^{2}d(d-1)
f_{2}^{2}t^{2f_{2}}-4ABdf_{1}f_{2}t^{f_{1}+f_{2}}+
8Af_{1}t^{f_{1}}\left(-2+f_{1}\left(3A
t^{f_{1}}+2\right)\right)\right)\right)}{\left(24kt^{2}+
e^{2At^{f_{1}}}\left(24A^{2}f_{1}^{2}t^{2f_{1}}-B d f_{2}
t^{f_{2}}\left(f_{2}-4\left(4+B(d+1)
t^{f_{2}}\right)\right)\right)\right)}}}~dt
\end{equation}

\begin{figure}
\includegraphics[scale=0.75]{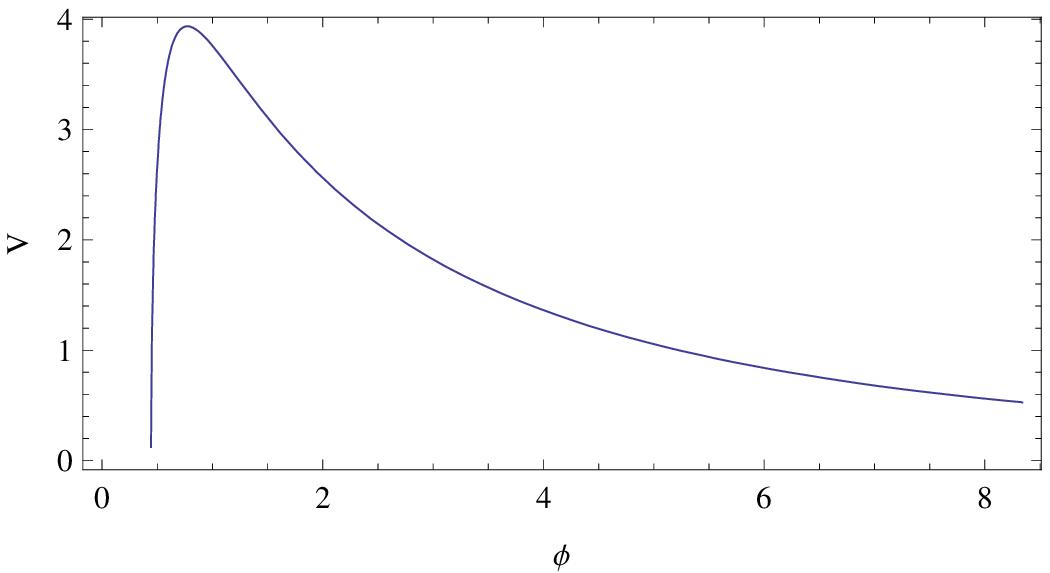}~~~~~~~~~~~~
\includegraphics[scale=0.45]{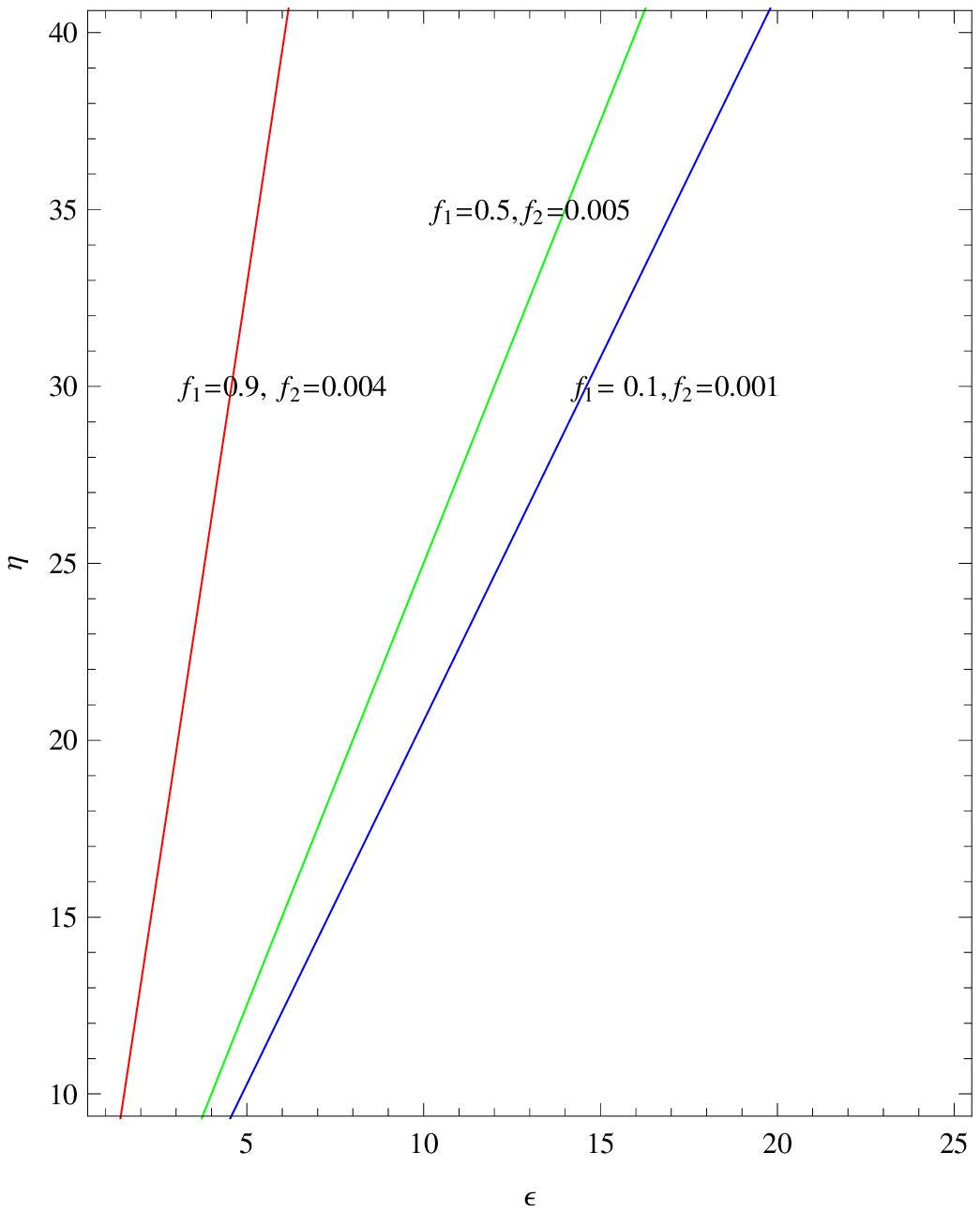}
\vspace{2mm}
~~~~~~~Fig.27~~~~~~~~~~~~~~~~~~~~~~~~~~~~~~~~~~~~~~~~~~~~~~~~~~~~~~~~Fig.28~~\\
\vspace{6mm} Fig. 27 shows the variations of $V$ against $\phi$,
for $A=0.5, B=0.02, k=1, f_{1}=0.5, f_{2}=0.2, d=5$ and Fig. 28
shows the variation of the slow roll parameters $\epsilon$ against
$\eta$ for $A=400, B=200, d=5$ and $k= -1, 1, 0$ respectively in
the case of Tachyonic Scalar field for Intermediate Scenario.
\vspace{6mm}
\end{figure}

$\bullet$ \textbf{Normal Scalar field:}\\

The energy density $\rho$ and pressure $p$ due to the Normal
Scalar field $\phi$ are given by the equations (20) and (21).
Using equations (55)-(57), we can find the expressions for
$V(\phi)$ and $\phi$ as

\begin{equation}
V(\phi)=\frac{1}{8t^{2}}e^{-2At^{f_{1}}}\left(16kt^{2} + e^{2A
t^{f_{1}}}\left(-B^{2}d^{2}f_{2}^{2}t^{2f_{2}}
-2Bdf_{2}t^{f_{2}}(Af_{1}t^{f_{1}}+f_{2}-1)+8Af_{1} t^{f_{1}} (3A
f_{1} t^{f_{1}}+ f_{1}-1)\right)\right)
\end{equation}
and
\begin{equation}
\phi=\int{\frac{1}{2}\sqrt{\left(\frac{1}{t^{2}}e^{-2At^{f_{1}}}
\left(8kt^{2}-e^{2At^{f_{1}}}\left(2Af_{1} t^{f_{1}}(-Bdf_{2}
t^{f_{2}}+4f_{1}-4)+ Bdf_{2} t^{f_{2}}(-2+
f_{2}(2+Bt^{f_{2}}))\right)\right)\right)}}~dt
\end{equation}

\begin{figure}
\includegraphics[scale=0.7]{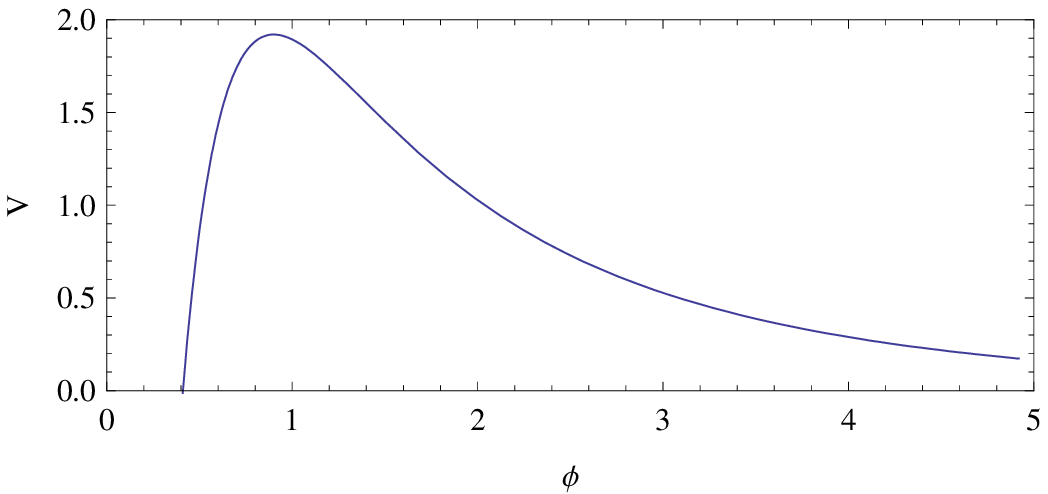}~~~~
\includegraphics[scale=0.55]{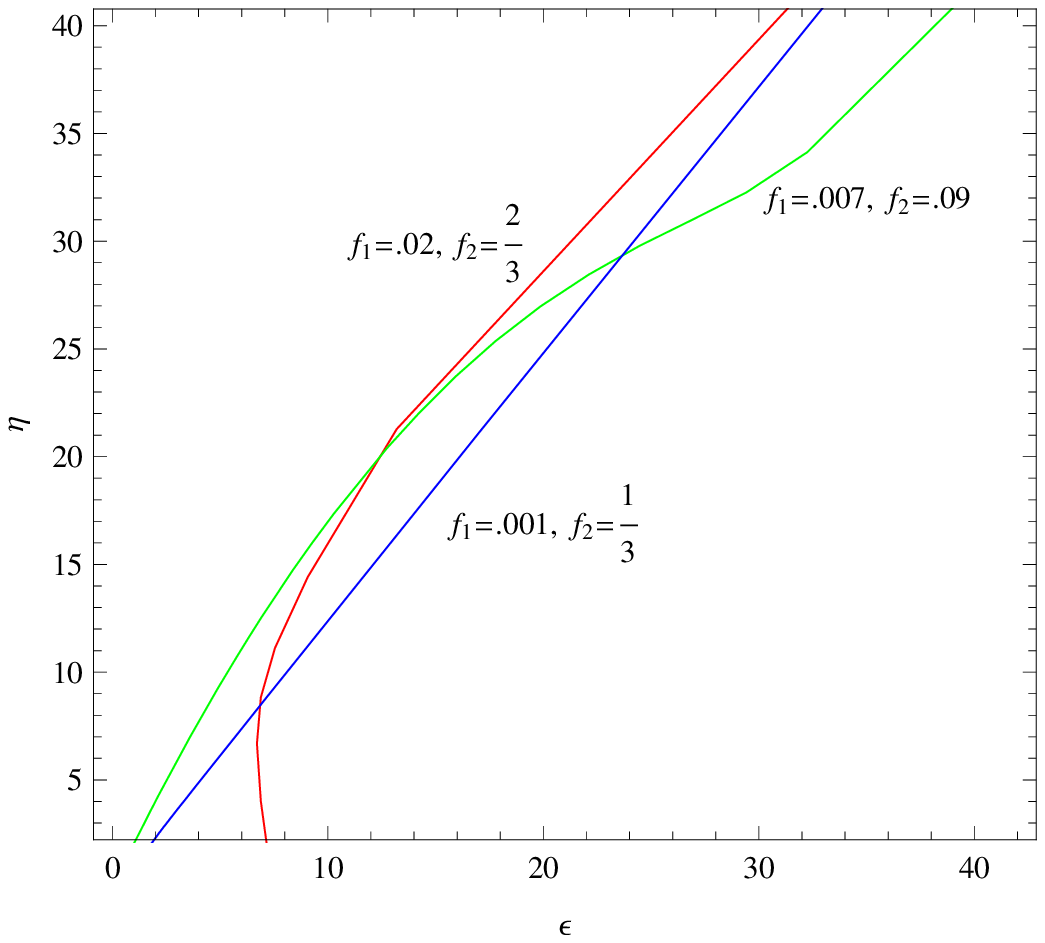}
\vspace{2mm}
~~~~Fig.29~~~~~~~~~~~~~~~~~~~~~~~~~~~~~~~~~~~~~~~~~~~~~~~~~~~~~~~~~~~~~~~~~~~~~Fig.30~~\\
\vspace{6mm} Fig. 29 shows the variations of $V$ against $\phi$,
for $A=1, B=0.1, k=1, f_{1}=0.3, f_{2}=0.4, d=5$ and Fig. 30 shows
the variation of the slow roll parameters $\epsilon$ against
$\eta$ for $A=.5, B=.3, d=5$ and $k= -1, 1, 0$ respectively in the
case of Normal Scalar field for Intermediate Scenario.
\vspace{6mm}
\end{figure}

$\bullet$ \textbf{DBI-essence:}\\

The energy density $\rho$ and pressure $p$ due to the DBI-essence
field $\phi$ are given by the equations (24) and (25).

{\bf Case I:}  $\gamma$ = constant.\\

Using equations (55)-(57), we can find the expressions for
$V(\phi)$, $T(\phi)$ and $\phi$ as

\begin{equation}
V(\phi)=\delta \phi_{0}^{2} e^{(2AFt^{f_{1}}+2BGt^{f_{2}})}
\end{equation}
\begin{equation}
T(\phi)=\sigma \phi_{0}^{2} e^{(2AFt^{f_{1}}+2BGt^{f_{2}})}
\end{equation}
and
\begin{equation}
\phi=\int{\phi_{0} e^{(AFt^{f_{1}}+BGt^{f_{2}})}}~dt
\end{equation}

where$E=\frac{1}{\gamma}[2(\delta-\sigma)+2(\sigma-1)\gamma^{3}]$,
$\gamma=\sqrt{\frac{\sigma}{\sigma-1}}$, $F=-\frac{3}{E}$,
$G=-\frac{d}{E}$ and $\phi_{0}$ is an integrating constants. From
above, we see that $V(\phi)$ and $T(\phi)$ are both exponentially
decreasing with DBI scalar field
$\phi$.\\

\begin{figure}
\includegraphics[scale=0.6]{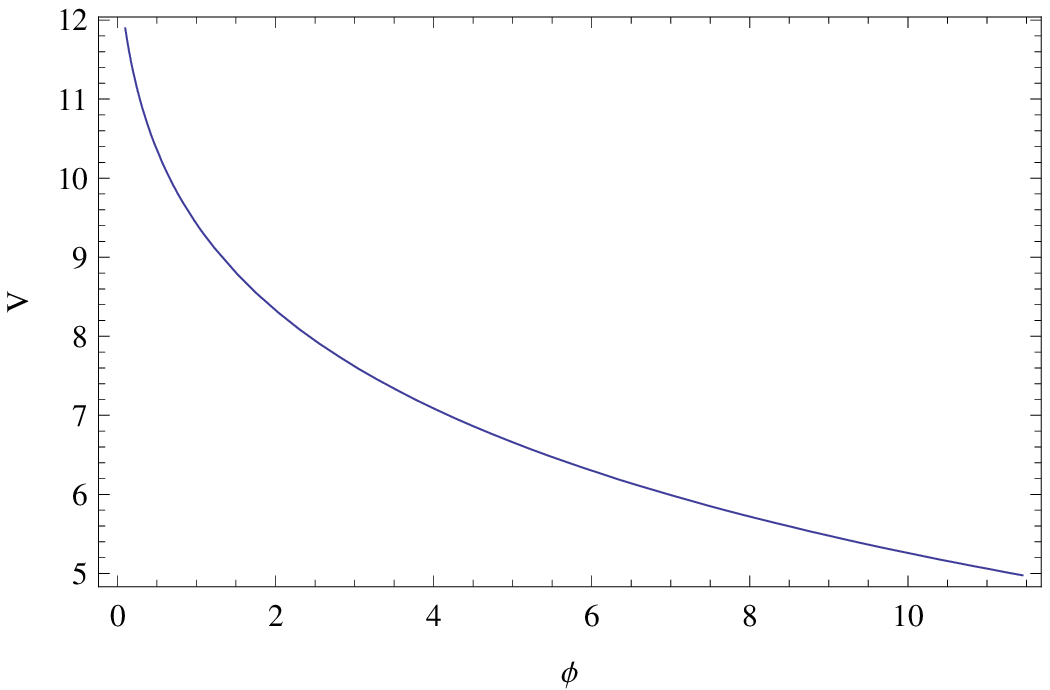}~~~~~~~~~~~~~~
\includegraphics[scale=0.65]{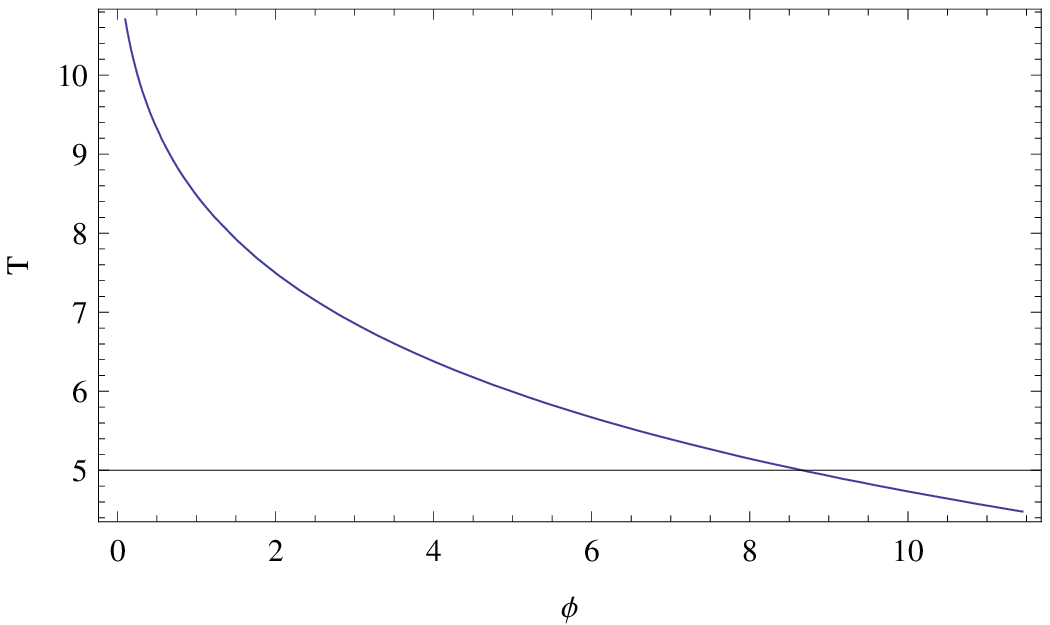}\\
\vspace{2mm}
~~~~~~~Fig.31~~~~~~~~~~~~~~~~~~~~~~~~~~~~~~~~~~~~~~~~~~~~~~~~~~~~~~~~~~~~~~Fig.32~~~~~~\\
\vspace{4mm}
\includegraphics[scale=.35]{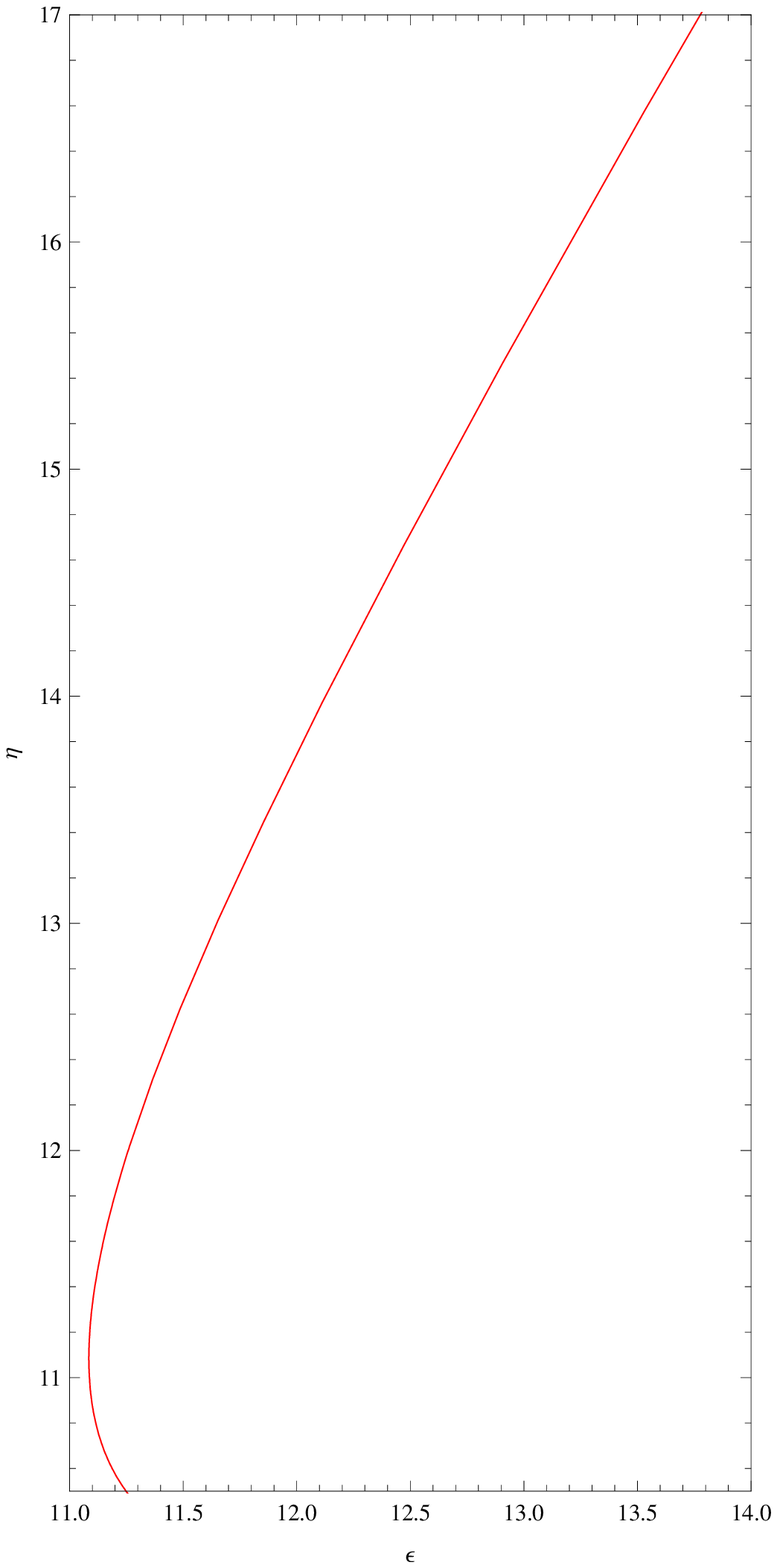}\\
\vspace{2mm}
~~~~~~~~~~~~~~~~~~~~~~~~~~~~~~~~~~~~~~~Fig.33~~~~~~~~~~~~~~~~~~~~~~~~~~~~~~~~\\
\vspace{6mm} Fig. 31 shows the variations of $V$ against $\phi$
and Fig. 32 shows the variations of $T$ against $\phi$, for $A=1,
B=1, f_{1}=0.4, f_{2}=0.1, \sigma=9, \delta=10, d=5, \phi_{0}=2$
and Fig. 33 shows the variation of the slow roll parameters
$\epsilon$ against $\eta$ for $A=1, B=2, \sigma=5,
\delta=2,f_{1}=0.1,f_{2}=0.2,\phi_{0}=2,d=5$ in the $1st$ case of
DBI-essence Scalar field for Intermediate Scenario.
 \vspace{6mm}
\end{figure}

{\bf Case II:}  $\gamma \neq$  constant.\\

Again using equations (55)-(57), we can find the expressions for
$V(\phi)$ and $\phi$ as

\begin{equation}
V(\phi)=\ln{[\frac{C_{0}}
{e^{(3At^{f_{1}}+dBt^{f_{2}})}}]}\sqrt{1-\frac{1}{\ln{[\frac{C_{0}}
{e^{(3At^{f_{1}}+dBt^{f_{2}})}}]}}}
\end{equation}
and
\begin{equation}
\phi=\int{\left(1-\frac{1}{\ln{[\frac{C_{0}}
{e^{(3At^{f_{1}}+dBt^{f_{2}})}}]}}\right)^{\frac{1}{4}}}~dt
\end{equation}

where $C_{0}$ is an integrating constant.\\\\

\begin{figure}
\includegraphics[scale=.4]{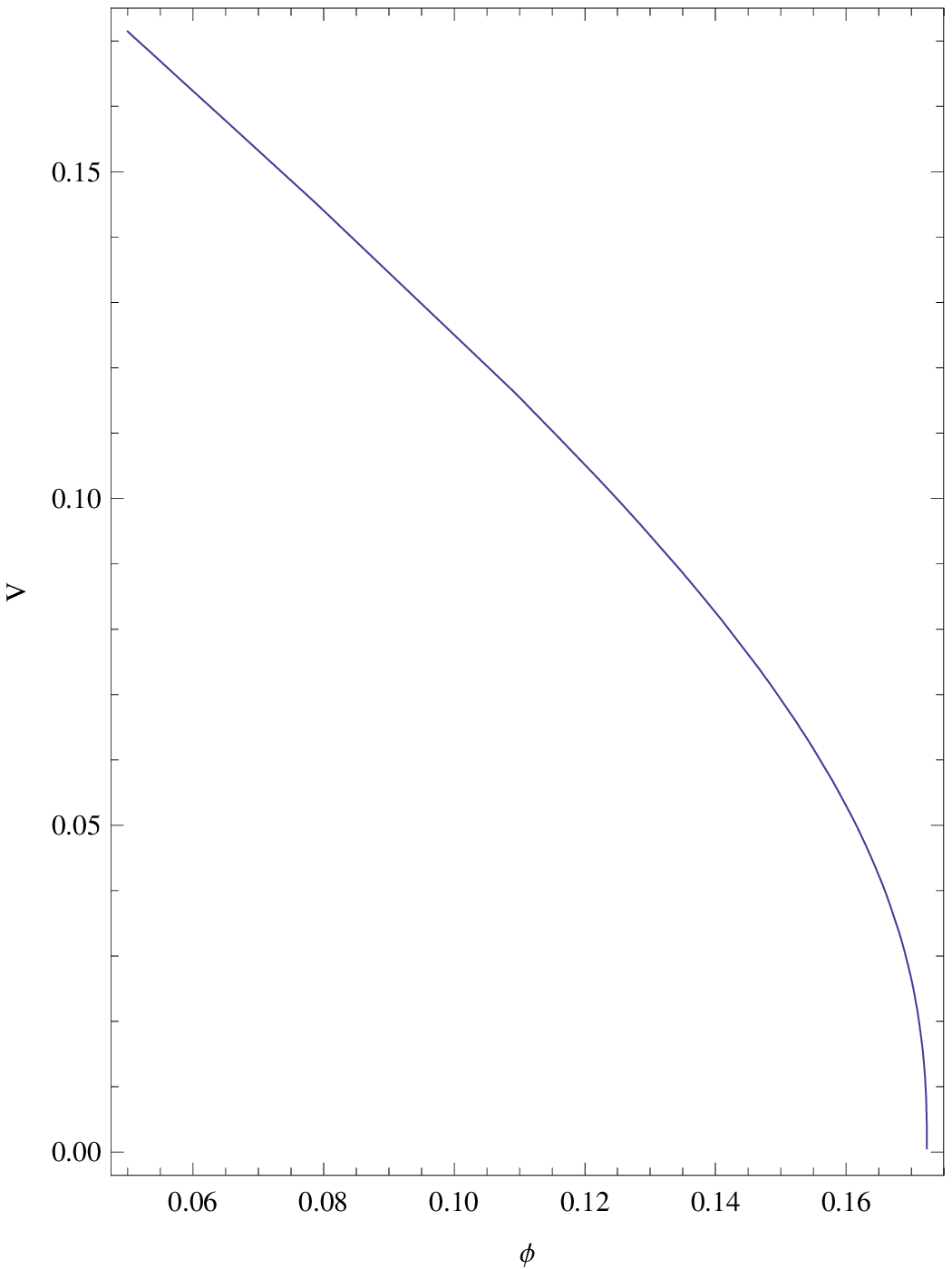}~~~~~~~~~~~~~~~~~~~~
\includegraphics[scale=.5]{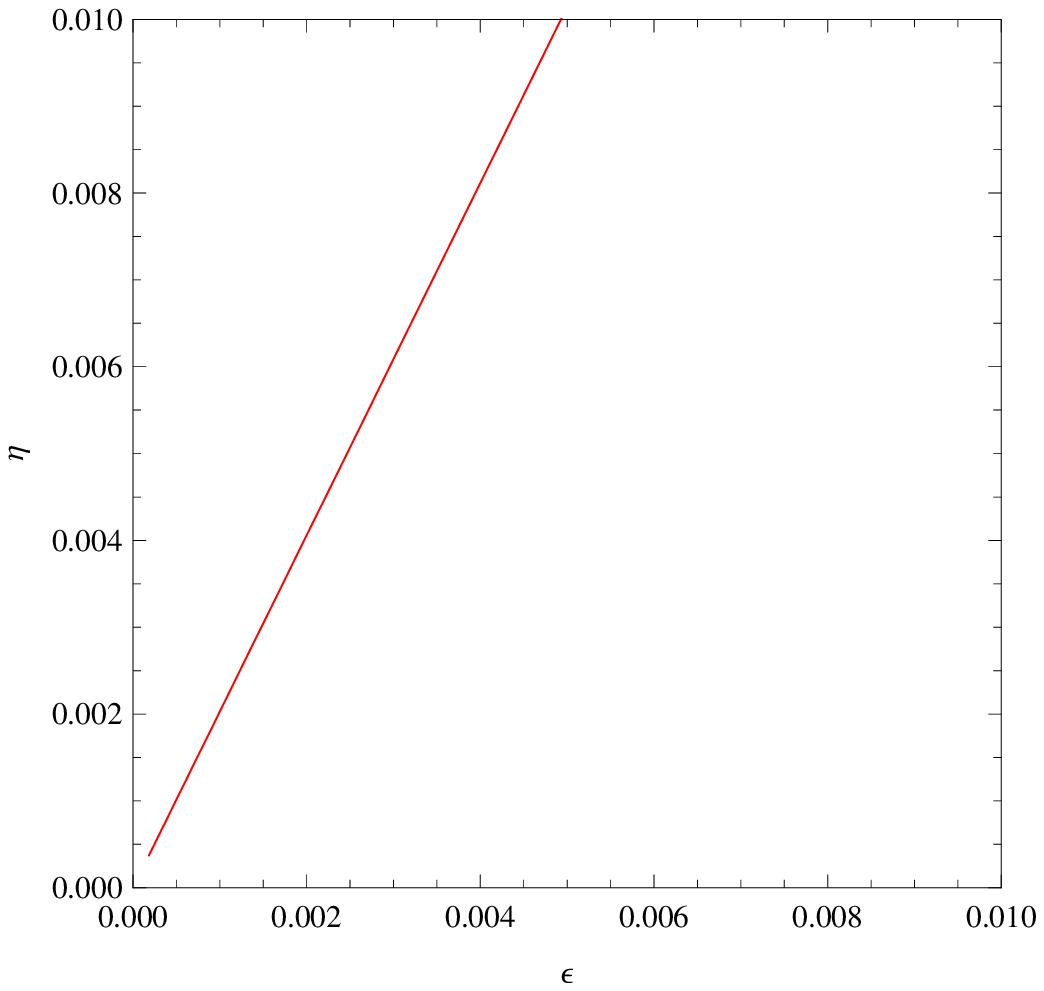}\\
\vspace{2mm}
~~~Fig.34~~~~~~~~~~~~~~~~~~~~~~~~~~~~~~~~~~~~~~~~~~~~~~~~~~~~~~~~~~~Fig.35~~~~~~\\
\vspace{6mm} Fig. 34 shows the variations of $V$ against $\phi$,
for $A=1, B=1, f_{1}=0.9, f_{2}=0.01,C_{0}=1000, d=5$  and Fig. 35
shows the variation of the slow roll parameters $\epsilon$ against
$\eta$ for $A=1, B=1, f_{1}=0.02, f_{2}=0.03,C_{0}=3, d=15$ in the
$2nd$ case of DBI-essence Scalar field for Intermediate Scenario.
\vspace{6mm}
\end{figure}

$\bullet$ \textbf{Statefinder parameters:}\\

The geometrical parameters\{$r,s$\} for higher dimensional
anisotropic cosmology in Intermediate scenario can be constructed
from the scale factors $a(t)$ and $b(t)$ as

\begin{equation}
r=1+\frac{3(d+3)(3A(f_{1}-1)f_{1}t^{f_{1}}+Bdf_{2}(f_{2}-1)t^{f_{2}})}{(3Af_{1}t^{f_{1}}+Bdf_{2}t^{f_{2}})^{2}}+
\frac{(3+d)^{2}(3Af_{1}(2+f_{1}(f_{1}-3))t^{f_{1}}+Bdf_{2}(2+f_{2}(f_{2}-3))t^{f_{2}})}{(3Af_{1}t^{f_{1}}+Bdf_{2}t^{f_{2}})^{3}}
\end{equation}
\begin{eqnarray*}
s=\frac{-\left.[(d+3)\left(3(3Af_{1}t^{f_{1}}+Bdf_{2}t^{f_{2}})(3A(f_{1}-1)f_{1}t^{f_{1}}+Bd
f_{2}(f_{2}-1)t^{f_{2}})
+(3+d)\left(3Af_{1}(2+f_{1}(f_{1}-3))t^{f_{1}}\right.\right.\right.}{[3(3Af_{1}t^{f_{1}}+
Bdf_{2}t^{f_{2}})^{3}(\frac{3}{2}+\frac{(d+3)(3A(f_{1}-1)f_{1}t^{f_{1}}+
Bd
f_{2}(f_{2}-1)t^{f_{2}})}{(3Af_{1}t^{f_{1}}+Bdf_{2}t^{f_{2}})^{2}})]}
\end{eqnarray*}
\begin{equation}
\frac{\left.\left.\left.+Bd
f_{2}(2+f_{2}(f_{2}-3))t^{f_{2}}\right)\right)\right]}{}
\end{equation}

\begin{figure}
\includegraphics[scale=0.6]{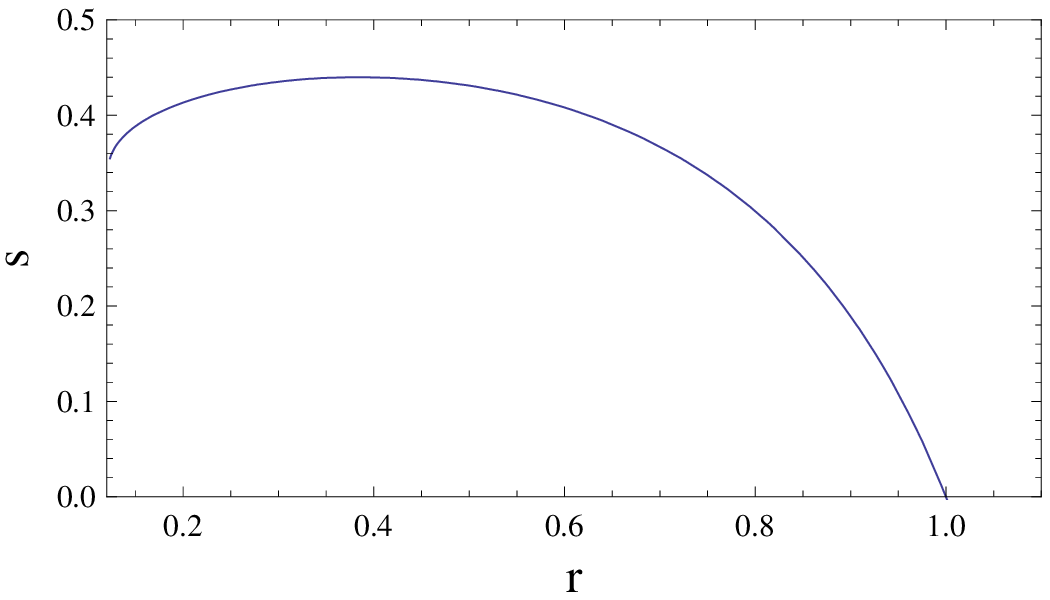}~~~~

\vspace{2mm}
~~~~~~~~~~~~~~~~~~Fig.36~~~~~~~~~~~~~~~~~~~\\
\vspace{6mm} Fig. 36 shows the variations of $r$ against $s$, for
$A=0.2, B=40, f_{1}=0.55, f_{2}=0.02, d =5 $in Intermediate
Scenario. \vspace{6mm}
\end{figure}

\section{\normalsize\bf{Discussions}}

In this work, we have considered $N~(=4+d)$-dimensional Einstein's
field equations in which 4-dimensional space-time is described by
a FRW metric and that of the extra $d$-dimensions by an Euclidean
metric. We have considered three scenarios, namely, Emergent,
Intermediate and Logamediate scenarios where the universe is
filled with K-essence, Tachyonic, Normal Scalar Field and
DBI-essence types dark energy models. The natures of the
potentials as well as dynamics of scalar fields for the dark
energy models have been analyzed. The statefinder and slow-roll
parameters have been considered and their natures have been
investigated for all dark energy models due to three scenarios of the universe.\\

In the case of Emergent scenario, we have considered a particular
forms of scale factors $a$ and $b$ in such a way that there is no
singularity for evolution of the anisotropic Universe. We have
found $\phi$ and potential $V$ in terms of cosmic time $t$ for
K-essence, Tachyonic, Normal Scalar Field and DBI-essence models.
Here we have shown that the emergent scenario is possible for
open, closed or flat Universe if the Universe contains K-essence,
Tachyonic, Normal Scalar Field and DBI-essence field. From figures
1, 3, 5, 7, 8, 10 it has been seen that the potential always
increases with K-essence, Tachyonic and decreases with Normal
Scalar Field and also with DBI-essence field when $\gamma=$
constant and $\gamma\neq$ constant and also the figures 2, 4, 6,
9, 11 shows the variation of slow-roll parameters $\epsilon$ and
$\eta$ in above dark energy model for open,closed and flat
universe where they are increase with all dark energy field except
Normal scalar field where it increases 1st then decreases. The
$\{r,s\}$ diagram (fig.12) shows that the evolution of the
emergent Universe starts from asymptotic Einstein's static era
($r\rightarrow \infty,~s\rightarrow -\infty$) and goes to
$\Lambda$CDM model ($r=1,~s=0$). It is also observed that $r,s$
are independent of the dimension $d$. So, from statefinder
parameters, the behavior of different stages of the evolution
of the emergent Universe have been generated.\\

In the case of Logamediate scenario, we have considered a
particular forms of scale factors $a$ and $b$ in such a way that
there is no singularity for evolution of the anisotropic Universe.
We have found $\phi$ and potential $V$ in terms of cosmic time $t$
for K-essence, Tachyonic, Normal Scalar Field and DBI-essence
models. Here we have shown that the logamediate scenario is
possible for open, closed or flat Universe if the Universe
contains K-essence, Tachyonic, Normal Scalar Field and DBI-essence
field. From figures 13, 15, 17, 19, 20, 22 it has been seen that
the potential are increases with K-essence, Tachyonic and
decreases with Normal Scalar Field and also with DBI-essence field
when $\gamma=$ constant and $\gamma\neq$ constant and also the
figures 14, 16, 18, 21, 23 shows the variation of slow-roll
parameters $\epsilon$ and $\eta$ in above dark energy models for
open, closed and flat universe where they are increasing with all
dark energy field except DBI essence scalar field where it
increases 1st then decreases then again increases. The $\{r,s\}$
diagram (fig.24) shows that the evolution of the Universe starts
from asymptotic Einstein static era ($r\rightarrow
\infty,~s\rightarrow -\infty$) and goes to $\Lambda$CDM model
($r=1,~s=0$). It is also observed that $r,s$ are independent of
the dimension $d$. So, from statefinder parameters, the behavior
of different stages of the evolution of the Logamediate Universe
have been generated.\\

In the case of Intermediate scenario, we have considered a
particular forms of scale factors $a$ and $b$ in such a way that
there is no singularity for evolution of the anisotropic Universe.
We have found $\phi$ and potential $V$ in terms of cosmic time $t$
for K-essence, Tachyonic, Normal Scalar Field and DBI-essence
models. Here we have shown that the intermediate scenario is
possible for open, closed or flat Universe if the Universe
contains K-essence, Tachyonic, Normal Scalar Field and DBI-essence
field. From figures 25, 27, 29, 31, 32, 34 it has been seen that
the potential are decreases with K-essence, DBI-essence field when
$\gamma=$ constant and $\gamma\neq$ constant and 1st increases
then decreases with Normal Scalar Field and with Tachyonic and
also the figures 26, 28, 30, 33, 35 shows the variation of
slow-roll parameters $\epsilon$ and $\eta$ in above dark energy
model for open,closed and flat universe where they are increasing
with all dark energy field except DBI essence scalar field where
it increases 1st then decreases then again increases. The
$\{r,s\}$ diagram (fig.36) shows that the evolution of the
Universe starts from asymptotic Einstein static era ($r\rightarrow
\infty,~s\rightarrow -\infty$) and goes to $\Lambda$CDM model
($r=1,~s=0$). It is also observed that $r,s$ are independent of
the dimension $d$. So, from statefinder parameters, the behavior
of different stages of the evolution of the Intermediate Universe
have been generated.\\

\textbf{References: }\\
\\
$[1]$ A.G.Riess et al, {\it Astrophys. J.}, {\bf 116}:1009-1038 (1998).\\
$[2]$ S. Perlmutter et al, {\it Astrophys. J.}, {\bf 517}:565-586 (1999).\\
$[3]$ U. Seljak et al [S D S S collaboration] \textit{Phys. Rev.},{\bf D71}:103515 (2005).\\
$[4]$ D.N.Spergel et al[W M A P collaboration],{\it Astrophys. J.Suppl.}, {\bf 170}:337 (2007).\\
$[5]$ A.G.Riess et al {\it Astrophys. J.}, {\bf 659},(2007),\\
$[6]$ D.J.Eisenstein et al[S D S S collaboration]{\it Astrophys. J.},{\bf633}:560-574(2005).\\
$[7]$ G.Cognola et al \textit{Phys. Rev. D} \textbf{79}:044001 (2009).\\
$[8]$ S.Chakraborty and U. Debnath, {\it Int. J. Theor. Phys.} {\bf 24}: 25 (2010).\\
$[9]$ C. Brans and R.H. Dicke, \textit{Phys. Rev.} \textbf{124}:925(1961).\\
$[10]$ R. Maartens,Reference Frames and Gravitomagnetism, ed. J Pascual-Sanchez et al. (World Sci., 2001), p93-119.\\
$[11]$ R.R.Caldwell, \textit{Phys. Lett. B} {\bf 545}:23 (2002).\\
$[12]$  A. Sen, \textit{JHEP} \textbf{0207}: 065 (2002).\\
$[13]$ H.Wei et al {\it class. Quantum. Grav.} {\bf 22}:3189(2005).\\
$[14]$ M. Susperregi {\it Phys. Rev. D} {\bf 68}:103509(2003).\\
$[15]$  C. Armendariz - Picon, V. F. Mukhanov and P. J. Steinhardt,  \textit{Phys. Rev. Lett.} \textbf{85}: 4438 (2000).\\
$[16]$ M.Spalinski \textit{JCAP},{\bf 05}:017(2007).\\
$[17]$ T. Kaluza, \textit{Preus. Acad. Wiss} \textbf{F1}: 9669(1921).\\
$[18]$ O. Klein, \textit{A. Phys} \textbf{37}: 895 (1926).\\
$[19]$ L. Randall and R. Sundrum, \textit{Phys. Rev. Lett}\textbf{83}: 3370 (1999).\\
$[20]$ P. S. Wesson, (1999) \textit{Space-Time-Matter},  World Scientific.\\
$[21]$ G.F.R.Ellis, R. Maartens {\it Class Quantum Gravity} {\bf 21}: 223 (2004).\\
$[22]$ G.F.R.Ellis, J. Murugan and C.G. Tsagas, {\it Class. Quant. Grav.} {\bf 21} :233 (2004).\\
$[23]$ S. Mukherjee, B. C. Paul, N. Dadhich, S. D. Maharaj, A. Beesham {\it Class. Quam. Grav.}{\bf 23}:6927(2006).\\
$[24]$ J. D. Barrow and N. J. Nunes \textit{Phys. Rev. D}, \textbf{76}: 043501 (2007).\\\
$[25]$ J. D. Barrow, {\it Phys. Lett. B} {\bf 235}: 40 (1990).\\
$[26]$ J. D. Barrow and P. Saich, {\it Phys. Lett. B} {\bf 249}:406 (1990).\\
$[27]$ J. D. Barrow, A. R. Liddle, and C. Pahud, {\it Phys. Rev.D} {\bf 74}:127305 (2006).\\\
$[28]$ V. Sahni, T. D. Saini, A. A. Starobinsky and U. Alam, {\it JETP Lett.} {\bf 77}:201 (2003).\\
$[29]$ B. C. Paul, {\it Phys. Rev. D} {\bf 64}:027302 (2001).\\
$[30]$ I. Pahwa, D. Choudhury and T. R. Seshadri, arXiv:1104.1925v1 [gr-qc].\\
$[31]$ U. Debnath: {\it Class. Quant. Grav.} \textbf{25}:205019 (2008)\\
$[32]$ S. Chakraborty and U. Debnath, {\it Int. J. Theor. Phys.}{\bf 50}:80 (2011).\\
$[33]$ J. Martin, M. Yamaguchi, Phys. Rev. {\bf D77}:123508(2008).\\
$[34]$ U. Debnath and M. Jamil, arXiv:1102.1632v1 [physics.gen-ph].\\
$[35]$ B.Gumjudpai, Ward, J.: Phys. Rev. {\bf D80}:023528 (2009).\\
$[36]$ J.Martin, Yamaguchi, M.: Phys. Rev. {\bf D77}:123508 (2008).\\
$[37]$ S.Chakraborty and U. Debnath, {\it Int. J. Theor. Phys.} {\bf 49}: 1693-1698 (2010).\\

\end{document}